\documentclass[sigconf,nonacm]{acmart}
\renewcommand\footnotetextcopyrightpermission[1]{}
\usepackage{savesym}
\usepackage[T1]{fontenc}
\restoresymbol{TXF}{iint}
\usepackage{amsfonts}
\usepackage{textcomp}
\usepackage{bbding}
\usepackage{pifont}
\usepackage[normalem]{ulem}
\usepackage{hyphenat}
\usepackage{mathtools}
\usepackage{xspace}
\usepackage{enumitem}
\usepackage{multirow}
\usepackage{float}
\usepackage{placeins}
\usepackage{booktabs}
\usepackage{tabularx}
\usepackage{makecell}
\usepackage{color}
\usepackage{xcolor}
\usepackage{tikz}
\usepackage{pgfplots}
\usetikzlibrary{patterns}
\usetikzlibrary{positioning}
\usepackage{amsmath}
\usepackage{multicol}
\usepackage{titlesec}
\usepackage[english]{babel}
\usepackage[utf8]{inputenc}
\usepackage{algorithm}
\usepackage[noend]{algpseudocode}
\usepackage{todonotes}
\pgfplotsset{compat=1.18}

\titleformat*{\section}{\Large\bfseries\MakeUppercase}

\titlespacing*{\section}{0pt}{4pt plus 2pt minus 1pt}{4pt plus 2pt minus 2pt}
\titlespacing*{\subsection}{0pt}{2pt plus 1pt minus 0pt}{2pt plus 1pt minus 0pt}
\algblockdefx[Upon]{Upon}{EndUpon}[1]{\textbf{Upon} #1}{}
\algtext*{EndUpon}
\DeclareMathOperator*{\argmax}{arg\,max}

\def\sys{AutoPilot\xspace}

\begin{document}

\fancyhead{}

\title{\sys: Learning to Steer High Speed Robust BFT}

\author{Liangrong Chen}
\affiliation{%
  \institution{City University of Hong Kong}}
\email{lchen653@cityu.edu.hk}

\author{Yue Zhang}
\affiliation{%
  \institution{New York University Courant}}
\email{yz11727@nyu.edu}

\author{Eric Zhou}
\affiliation{%
  \institution{Troy High School}}
\email{erichyzhou@gmail.com}

\author{Mohammad Javad Amiri}
\affiliation{%
  \institution{Stony Brook University}}
\email{amiri@cs.stonybrook.edu}

\author{Ryan Marcus}
\affiliation{%
  \institution{University of Pennsylvania}}
\email{rcmarcus@seas.upenn.edu}

\author{Chenyuan Wu}
\affiliation{%
  \institution{City University of Hong Kong}}
\email{chenyuan.wu@cityu.edu.hk}

\begin{abstract}
Recent Byzantine Fault Tolerant (BFT) protocols achieve strong performance by combining the low-latency advantages of leader-based BFT protocols with the high-throughput benefits of DAG-based data dissemination. Despite exposing a wide spectrum of internal tunable parameters, these protocols typically rely on static and heuristic configurations, which leads to performance degradation under dynamic workloads, heterogeneous network conditions, and evolving adversarial behaviors.
In this paper, we present \sys, a reinforcement learning-based framework that continuously monitors runtime conditions and dynamically adjusts protocol parameters online to optimize consensus performance. To ensure robustness, \sys coordinates learning in a decentralized manner, providing resilience against adversarial data pollution. We implement \sys on top of Autobahn, a state-of-the-art, high-speed, robust BFT protocol, and evaluate it across diverse dynamic environments. Experimental results demonstrate that \sys quickly converges to the optimal configuration under changing environments, reduces end-to-end latency by 49.8\% compared to the default protocol configuration, and outperforms random configuration exploration by 73.3\%.
\end{abstract}

\maketitle
\section{Introduction}\label{sec:intro}

Byzantine Fault Tolerant (BFT) protocols are critical building blocks in untrustworthy distributed data management systems. By tolerating up to $f$ Byzantine adversaries, these protocols guarantee strong consistency through agreement on a common transaction order across replicas. BFT protocols have been widely deployed in distributed applications such as permissioned blockchains~\cite{amiri2019caper, HyperledgerUrsa, buchnik2020fireledger, resilientdb, amiri2021sharper}, permissionless blockchains~\cite{brown2016corda, kogias2016enhancing, kokoris2018omniledger, luu2016secure}, confidential consortium frameworks~\cite{microsoftCCF}, distributed file systems~\cite{adya2002farsite, castro2002practical, clement2009upright}, locking services~\cite{clement2009making}, firewalls~\cite{bessani2013depsky, garcia2016sieveq, roeder2010proactive, sousa2009highly}, certificate authority systems~\cite{zhou2002coca}, SCADA systems~\cite{babay2019deploying, kieckhafer1994reaching}, key-value datastores~\cite{dobre2013powerstore, goodson2004efficient} and key management~\cite{malkhi1998secure}.

Traditional view-based BFT protocols~\cite{castro1999practical,amir2011prime,gueta2019sbft,kotla2007zyzzyva,yin2019hotstuff} are optimized for responsiveness during fault-free, synchronous periods. However, they suffer from degraded performance when failures or network blips interrupt progress.
This degradation persists even after the failure is resolved, as backlogged requests accumulate and delay subsequent transactions. Directed Acyclic Graph (DAG)-based BFT protocols~\cite{danezis2022narwhal,spiegelman2022bullshark,keidar2021all, spiegelman2024shoal, malkhi2024bbca, shrestha2024sailfish, dai2024lightdag, cheng2024shardag, dai2024remora, dai2024wahoo, dai2023gradeddag,nagda2026dag,kang2025fairdag} alleviate this problem by decoupling data dissemination (i.e., building the DAG) from consensus (i.e., ordering the DAG) and propagating transactions asynchronously across replicas. This design enables high throughput and allows the system to continue making progress even during network disruptions. DAG-based BFT protocols have consequently been widely adopted in practical decentralized systems, such as Aptos~\cite{Aptos}, Sui~\cite{Sui}, Fantom~\cite{fantom}, and Avalanche~\cite{Avalanche}. However, their ordering mechanisms typically require multiple structured rounds and introduce non-trivial latency overhead.

State-of-the-art BFT protocols, like Autobahn~\cite{giridharan2024autobahn}, address both limitations simultaneously by combining a DAG-based asynchronous data dissemination layer with a low-latency, partially synchronous external consensus mechanism. In particular, Autobahn avoids the post-failure performance degradation of traditional BFT protocols, matches the high throughput of state-of-the-art DAG-based protocols, and achieves the low latency of traditional BFT protocols. Autobahn has been deployed in the latest production blockchains, including Sei Giga~\cite{seigiga}, Stable~\cite{stable}, and Somnia~\cite{Somnia}.

Despite these advances, Autobahn and other existing DAG-based BFT protocols face two major challenges. First, these protocols expose a wide spectrum of tunable configuration parameters, such as header size, round progression timeouts, and number of concurrent DAG instances~\cite{spiegelman2022bullshark, arun2025shoal++}. Optimally configuring these parameters is challenging due to the large parameter space and the intricate interactions among them. Second, these protocols are deployed in dynamic decentralized environments where workloads, network conditions and fault scenarios evolve, rendering any \emph{static} configuration quickly suboptimal. For instance, raising the number of concurrent DAG instances $d$ can reduce expected queuing latency by allowing transactions to be proposed more quickly; yet it also increases metadata overhead per round, especially under large network sizes. Consequently, no single value of $d$ consistently dominates across all operating conditions.

These observations suggest the need for an \emph{adaptive} mechanism that can continuously adjust protocol configurations at runtime in response to changing environments. Recent studies~\cite{li2023auto,li2025flexim,bahsoun2015making,chacko2023optimize,wu2025bftbrain,wu2022adachain,ding2026alzo,an2024rlchain,wu2024towards} have shown the feasibility of machine learning-based approaches for optimizing distributed protocols. For instance, BFTBrain~\cite{wu2025bftbrain} demonstrates that learned policies can effectively adapt BFT protocol behavior to dynamic conditions, achieving significant performance improvements over static designs. However, existing approaches primarily target traditional BFT protocols and do not address the unique characteristics and optimization challenges of DAG-based BFT systems. Moreover, they focus on inter-protocol switching rather than fine-grained intra-protocol parameter tuning.

To fill this gap, we propose {\em \sys}, a reinforcement learning (RL)-based framework that automatically tunes protocol parameters at runtime. At a high level, given a performance metric to optimize, \sys intelligently selects among a set of parameter configurations in response to dynamic changes in workload, network conditions, and fault scenarios. \sys brings three key operational benefits. First, \sys adapts to changing environments, eliminating the need for manual parameter selection across a vast configuration space. Second, \sys operates on a live system and reconfigures itself in real time, without a prolonged offline data collection phase prior to deployment. Third, \sys performs this adaptation in a decentralized, Byzantine fault-tolerant manner, without relying on a centralized machine learning agent.

To apply RL to this problem, \sys collects performance metrics from all nodes in a distributed manner. Nodes share locally observed features and reward signals through a coordination protocol, ensuring that all benign nodes converge to the same learning output and maintain resilience even when Byzantine nodes equivocate or forge observations. These metrics serve as features for \sys's RL engine, which models the selection of parameter configurations as a contextual multi-armed bandit (CMAB) problem~\cite{bandit_survey} and strategically switches among configurations at runtime to identify those best suited to current system conditions. Our evaluation demonstrates that \sys significantly outperforms static parameter configurations under dynamic networks, workloads, and fault scenarios. Specifically, this paper makes the following contributions:

\begin{itemize}[parsep=1mm, leftmargin=1em,labelwidth=*,align=left]
    \item {\bf Learned adaptive DAG-based BFT protocols.} \sys is the first system that \textit{learns} to automatically tune fine-grained parameters across both the DAG-based data dissemination layer and the external BFT consensus layer in hybrid BFT protocols such as Autobahn. Without requiring offline profiling or data collection prior to deployment, \sys enables automatic adaptation to unforeseen system conditions at runtime.

    \item {\bf Analysis of fine-grained parameters in DAG-based BFT protocols.} We conducted systematic experiments across a wide range of tunable parameters in Autobahn, providing insights into how parameter choices should adapt to a changing environment and how parameter interactions affect performance. Our results highlight the breadth of the state and action space, demonstrating that manual tuning is both inefficient and impractical.

    \item{\bf Prototype and experimental evaluation.} We developed a prototype of \sys and evaluated it on Google Cloud Platform. Experimentally, \sys reduces latency by 49.8\% compared to the default configuration under dynamic conditions, and outperforms the baseline without robust learning coordination by 27.6\%--281.8\% under data pollution attacks.
\end{itemize}
\section{Background} \label{sec:background}

\begin{figure}[t]
    \centering
    \includegraphics[width= 0.5\textwidth]{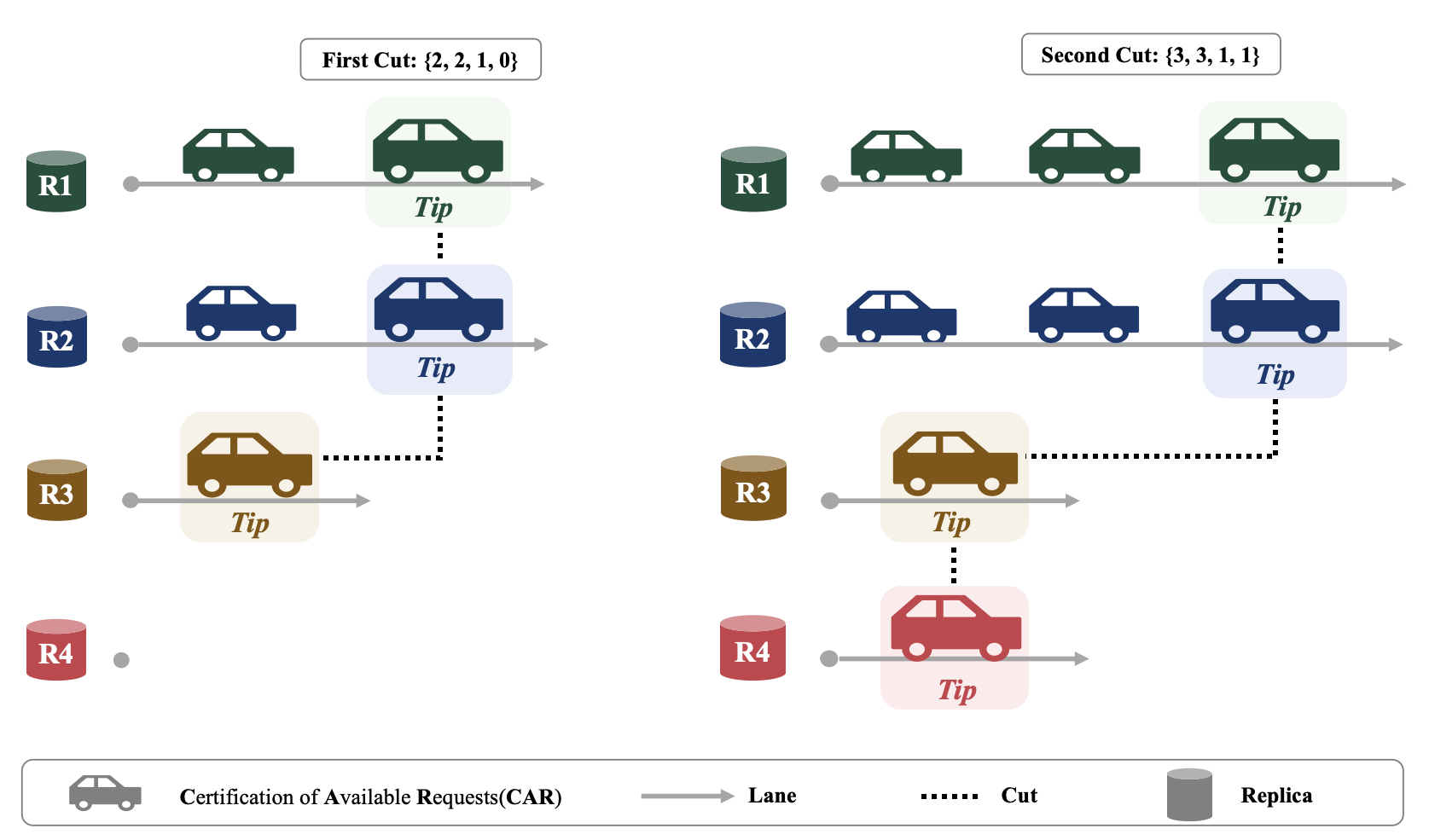}
    \caption{Example on Autobahn's {\em cuts}. When {\em cut condition} is configured to $2f+1$, a cut will be created once every $3$ new tips are generated.}
    \label{fig:lane_cut}
\end{figure}

Autobahn is a novel partially synchronous BFT consensus protocol that offers high throughput, low latency, and performance robustness. The data dissemination layer in Autobahn loosely resembles the DAG structure. Specifically, each replica maintains a chain of certified records known as a {\em lane}. To extend its lane, a replica starts by creating a {\em header} containing a batch of transactions together with the certificate of the latest certified record in its own lane. The replica then broadcasts the header to all other replicas. Once the proposer collects $f+1$ votes from distinct replicas, it aggregates the votes with the header to form a {\em CAR} (Certified Availability Record). The resulting CAR becomes the latest vertex (i.e., the certified record) in the replica's lane, and the proposer can proceed to create the next header. A replica votes for another replica's new header if and only if it has already voted for that header's parent, a rule known as {\em in-order voting}. This mechanism ensures that the complete history of each lane is possessed by at least $f+1$ replicas and can be fetched at once, thereby reducing data synchronization latency. The latest CAR in a lane is referred to as the {\em tip}. The collection of tips across all replicas forms a {\em cut}, which represents the current frontier of the DAG. 

Autobahn orders transactions through a sequence of consensus slots. Each slot $s$ is assigned to a leader, which becomes responsible for proposing it once the previous slot $s-1$ commits. Within each slot, Autobahn follows a classical leader-based consensus~\cite{castro1999practical}. To propose a new slot, the leader continuously monitors its local view of DAG and waits until the configured \textit{cut condition} is satisfied, i.e., enough new {\em tips} have been collected. Upon satisfying the cut condition, the leader creates a new {\em lane cut}, as illustrated in Figure~\ref{fig:lane_cut}. All blocks between the new {\em lane cut} and the previous {\em lane cut} are then aggregated into a proposal for a new slot. If the current leader fails to make timely progress, replicas trigger a view change and elect a new leader to continue the consensus process. By repeating the above process, Autobahn eventually generates a consistent linear sequence of transaction data across all validators.
\section{Why DAG consensus needs learning}\label{sec:motivation}

\begin{table*}[!t]
\centering
\caption{Scenarios representing various environmental conditions}
\label{tab:state_setting}
\small
\begin{tabular}{c>{\centering\arraybackslash}p{7.7cm}cc}
\toprule
\textbf{Scenario} & \textbf{Node Geographical Distribution} & \textbf{Affected Nodes, Duration, Change } & \textbf{Adversarial Behavior} \\
\midrule
$S_1$ & 2 in \texttt{Asia-east} and 2 in \texttt{US-central} & -- & -- \\
$S_2$ & 4 in Asia (2 in \texttt{Asia-southeast} and 2 in \texttt{Asia-east})  & -- & -- \\
$S_3$ & 2 in \texttt{Asia-east}, 1 in \texttt{Europe-central}, and 1 in \texttt{ME-central} & -- & -- \\
$S_4$ & 10 in Asia (All in \texttt{Asia-east}) & 5, [0, 120], 10 $\times$ workload increase& -- \\
$S_5$ & 4 in Asia (All in \texttt{Asia-east})  &  1, [0, 120], 200ms delay & Intentional Vote Delay\\
$S_6$ & 4 in Asia (All in \texttt{Asia-east})  & -- & Non-responsive \\
\bottomrule
\end{tabular}
\end{table*}

\begin{table*}[!t]
\centering
\begin{minipage}[t]{0.47\textwidth}
\centering
\caption{Parameter configurations}
\label{tab:action_setting}
\small
\begin{tabular}{@{}l>{\centering\arraybackslash}p{1.15cm}>{\centering\arraybackslash}p{0.85cm}ccc@{}}
\toprule
\textbf{Action} & \makecell{\textbf{FP}\\\textbf{Timeout}} &\makecell{\textbf{Cut}\\\textbf{Cond.}} & \makecell{\textbf{Parallel}\\\textbf{Proposal}} & \makecell{\textbf{Header}\\\textbf{Size}} & \makecell{\textbf{Max Header}\\\textbf{/Batch Delay}} \\
\midrule
$A_1$  & 100ms  & 2f+1 & 1 & 32B & 5000ms \\
$A_2$  & 100ms  & 2f+1 & 4 & 32B & 5000ms \\
$A_3$  & 500ms  & 2f+1 & 4 & 32B & 5000ms \\
$A_4$  & 0ms  & 2f+1 & 4 & 32B & 5000ms\\
$A_5$  & 100ms  & $ \lceil\frac{3f+1}{2}\rceil $ & 4 & 32B & 5000ms\\
$A_6$  & 0ms  & 2f+1 & 4 & 128B & 5000ms \\
\midrule
{\sf DEF} & 200ms & 2f+1 & 4 & 32B & 200ms\\
\bottomrule
\end{tabular}
\end{minipage}
\hfill
\begin{minipage}[t]{0.52\textwidth}
\centering
\caption{Latency comparison (increase compared to the best config.)}
\label{tab:performance_comparison}
\small
\begin{tabular}{clll}
\toprule
\textbf{Scenario} & \textbf{Best Config.} & \textbf{Default Config.} & \textbf{2nd Best Config.}  \\
\midrule
$S_1$ & $A_2$, 658ms  & 748ms  (\textbf{+13.6\%}) & {\sf DEF}  \\
$S_2$ & $A_1$, 158ms  & 532ms  (\textbf{+236.8\%}) & $A_6$, 219ms  (\textbf{+38.6\%})  \\
$S_3$ & $A_3$, 1143ms  & 1425ms (\textbf{+24.6\%}) & $A_2$, 1398ms (\textbf{+22.3\%})  \\
$S_4$ & $A_5$, 815ms   & 1232ms (\textbf{+51.2\%}) & $A_4$, 1224ms (\textbf{+50.1\%})   \\
$S_5$ & $A_4$, 60ms    & 248ms  (\textbf{+313.3\%}) & $A_2$, 144ms  (\textbf{+140.0\%}) \\
$S_6$ & $A_6$, 622ms   & 1262ms (\textbf{+102.9\%}) & $A_1$, 801ms  (\textbf{+28.8\%})  \\
\bottomrule
\end{tabular}
\end{minipage}
\end{table*}

While DAG-based BFT protocols offer greater scalability than traditional leader-based protocols, they also introduce more internal parameters for configuration. In this section, we examine how these knobs affect system performance under different environments. Our experiments are conducted atop Autobahn~\cite{giridharan2024autobahn}, a representative DAG-based BFT protocol widely deployed in practice~\cite{seigiga,stable,Somnia}. As illustrated in Table~\ref{tab:state_setting}, we evaluate its performance across six scenarios ($S_1$–$S_6$) that represent distinct runtime environment conditions, varying across three dimensions: \emph{node geographical distribution}, \emph{workload imbalance across replicas}, and \emph{adversarial behavior}; all of which could change over time in realistic deployments.

Table~\ref{tab:action_setting} defines the set of configurations that we evaluate in our experiment, including $A_1$–$A_6$ and a default baseline ({\sf DEF}) used by the Autobahn repository. These configurations vary across five key parameters: (1) \emph{Fast path timeout} controls how long the leader waits for $3f+1$ votes before falling back to the slow path, where a timeout value of $0$ indicates that the fast path is disabled; (2) \emph{Cut condition} determines the minimum number of lane tips required to initiate a new proposal; (3) \emph{Number of parallel proposals} decides how many consensus slots are allowed to pipeline concurrently; (4) \emph{Header/Batch size} sets the maximum number of transactions per batch and batches per header. (5) \emph{Max header/batch delay} bounds the waiting time for assembling a header or batch before it is forcibly emitted. We note that these parameters are also present in other protocols, like Narwhal~\cite{narwhalrepo2026} and HotStuff~\cite{kang2025hotstuff}.

We evaluate the configuration parameters listed in Table~\ref{tab:action_setting} under each scenario. Each experiment runs for $120$ seconds and is repeated twice; reported latency values are averaged across both runs.  Details about the experimental setup can be found in Section~\ref{sec:evaluation}.

Table~\ref{tab:performance_comparison} reports the best-performing and second-best configuration choices for each scenario, along with the default configuration. For each configuration, it reports the corresponding end-to-end latency and the percentage latency increase relative to the best-performing one. Importantly, it demonstrates that no single configuration works well across all scenarios, underscoring the need for adaptive parameter tuning. In the rest of this section, we explore why DAG-based BFT protocols require configuration changes as geography, workload, and adversarial behavior vary.

\subsection{Geographic Distribution}\label{subsec:geo_distribution}
Scenarios $S_1$, $S_2$, and $S_3$ demonstrate that both the proposal pipelining and the fast path mechanism are strongly influenced by the geographical distribution of replicas.

\noindent\textbf{Number of Parallel Proposals ($k$).} Normally, increasing $k$ improves performance by executing multiple proposals concurrently: with $k > 1$, up to $k-1$ slots can begin as soon as the prepare phase of the previous slot completes. For instance, in $S_1$, choosing action $A_1$ ($k=1$) leads to $28.1$\% higher latency compared to $A_2$ ($k=4$), demonstrating the benefit of pipelining. However, setting a larger $k$ is not always beneficial: in $S_2$, where consensus rounds are shorter than $S_1$ due to low RTTs, aggressive pipelining causes proposals to be generated more frequently, with each proposal carrying only a small number of blocks. This reduces the amortization of consensus costs across transactions while increasing communication overhead, ultimately degrading performance. In this case, $A_1$ ($k=1$) achieves the best performance. Increasing header size as in $A_6$ partially mitigates the overhead of frequent proposal initialization but still incurs $38.6$\% higher latency compared to $A_1$. Therefore, $k$ must be tuned carefully to balance the benefits of parallelism against the overhead of frequent proposal initialization.

\noindent\textbf{Fast Path Timeout.} Autobahn's linear consensus protocol follows the classic PBFT-style pattern, augmented with a fast path that reduces latency from the two-round slow path (5 message delays) to a single round (3 message delays). Specifically, upon collecting $2f+1$ matching {\small \sf Prepare} votes for a proposal, the leader waits for a \emph{fast path timeout} to determine whether $3f+1$ matching votes can be gathered. If so, the proposal is committed immediately via the fast path; otherwise, consensus falls back to the slow path and proceeds through an additional round of $2f+1$ quorum collections. This mechanism represents a fundamental trade-off between communication overhead and responsiveness.

In both $S_1$ and $S_2$, the network is organized into two clusters containing half of nodes respectively, such that intra-cluster communication is fast while cross-cluster communication is expensive. Since no single cluster has a sufficient number of nodes ($2f+1$) to establish consensus, the fast path typically requires one round of cross-cluster communication and collecting additional $f$ votes within the cluster, while the slow path needs to perform two rounds of cross-cluster communication. Under such conditions, taking the fast path is always beneficial.
Accordingly, $A_2$ and $A_1$ with a fast path timeout of 100\,ms achieve the best performance in $S_1$ and $S_2$ respectively, while disabling the fast path entirely, as in $A_4$ proves suboptimal in both scenarios. However, under slight network fluctuations where some slot fails to complete the fast path, the protocol is forced to wait for the full \emph{fast path timeout} before falling back to the slow path. As a result, an excessively long timeout, as in {\sf DEF} and $A_3$, directly increases the duration of this unnecessary waiting period, leaving more transactions blocked and increasing backlog, which in turn increases end-to-end latency.

In $S_3$, nodes are spread across three geographically distant regions, resulting in high RTTs. Here, the fast path remains beneficial, but
$A_2$'s shorter timeout leads to $22.3$\% higher latency compared to $A_3$'s longer timeout. The reason is that short timeout causes frequent fallbacks, as the system is unable to gather the additional $f+1$ responses from distant nodes before the timer expires. $A_3$'s extended timeout provides sufficient time to collect $3f+1$ responses, suggesting that the optimal fast-path timeout depends strongly on the underlying network latency and stability.

These results yield two key practical insights. First, in real-world deployments of DAG-based protocols such as public blockchains, nodes may join or leave dynamically, causing the network topology to change over time and necessitating continuous reconfiguration of protocol parameters. Second, the optimal \emph{fast path timeout} depends on prevailing global network latency, which may fluctuate due to network failures or adversarial interference; the timeout must therefore adapt dynamically to current network conditions.

\subsection{Imbalanced Workload}
In $S_4$, by scaling the system from $4$ to $10$ nodes and introducing five hotspot nodes that receive ten times more transactions than the remaining validators, we explore how workload imbalance across replicas affects the data dissemination and the initialization of proposals in the consensus layer.

\noindent\textbf{Cut Condition.} The \emph{cut condition} determines when a new proposal can be initiated. In $S_4$, five nodes receive ten times more transactions than the others, causing their lanes to grow significantly faster and producing an imbalance in \emph{tip} generation rates. When the cut condition is set too high, consensus must wait for tips from slower lanes before performing a lane cut and initiating a new proposal, introducing unnecessary latency. Accordingly, by reducing the cut condition from $2f+1$ to $\lceil (3f+1)/2 \rceil$, $A_5$ allows proposal generation to proceed without waiting for progress from all slow lanes, reducing latency by $33.5\%$ (from $1224$ to $815$) compared with the second-best configuration. In summary, the optimal \emph{cut condition} must adapt dynamically to the number of fast lanes, minimizing unnecessary delays imposed by slower nodes.

\noindent\textbf{Header Size/Max Header Delay.}
Header/Batch Size and Max Header/Batch Delay control how quickly new lane tips are generated in the data dissemination layer. Under highly imbalanced workloads, consensus progress is often constrained by the slowest lanes rather than the fastest ones. Larger header sizes require slow lanes to accumulate more transactions before generating a new header, further delaying tip generation and making it harder to satisfy the cut condition. Accordingly, under $S_4$, configuration 
$A_6$ (3205ms) incurs $161.8\%$ higher latency compared to $A_4$ (1224ms) and $160.1\%$ higher latency compared to the default configuration (1232ms), which uses a much smaller Max Header/Batch Delay.

\noindent\textbf{Number of Parallel Proposals ($k$).}
The effectiveness of proposal pipelining also depends on the \emph{cut condition}. Although increasing $k$ allows multiple consensus instances to pipeline, a new proposal cannot be generated until the \emph{cut condition} is satisfied. Under highly imbalanced workloads, proposal generation is frequently delayed by slow lanes, preventing the system from fully utilizing the available parallelism. For instance, the performance differences between $A_1$ and $A_2$ in $S_4$ remain relatively small despite their different pipelining configurations. This observation highlights a significant interaction between dissemination-layer and consensus-layer parameters: increasing parallelism alone does not guarantee better performance if proposal generation is bottlenecked by insufficient lane progress. The effectiveness of parallelism, therefore, depends on the setting of \emph{cut condition}.

\subsection{Adversarial Behavior}

Beyond geographic distribution and workload imbalance, we examine the effect of adversarial behaviors in $S_5$ and $S_6$.

\noindent\textbf{Intentional Vote Delay.} In $S_5$, $f$ nodes deliberately delay their consensus votes by 200 ms. Since the fast path requires $3f+1$ prepare votes, $f$ Byzantine replicas can deliberately delay their votes to prevent the leader from collecting a full quorum in time. As a result, the leader is forced to wait until the fast-path timeout expires before falling back to the slow path, introducing unnecessary delay. As shown in Table~\ref{tab:performance_comparison}, under scenario $S_5$, disabling the fast path as in $A_4$ renders the best performance, whereas $A_2$ and the default incur $145.0$\% and $313.3$\% higher latency, respectively.

\noindent\textbf{Non-responsive Nodes.} In $S_6$, $f$ Byzantine nodes cease responding entirely, leaving the protocol to rely on the remaining $2f+1$ honest nodes for progress. Gathering $3f+1$ prepare votes is impossible when $f$ nodes are silent, causing all fast path waits and falling back to the slow path. Consequently, $A_6$ that disables the fast path entirely achieves the best performance.
The larger header size further helps amortize consensus overhead across more transactions, providing an additional performance benefit.
Additionally, since the non-responsive nodes cease proposing new blocks, any cut condition requiring more than $2f+1$ new tips becomes unsatisfiable, preventing new proposal initialization and potentially triggering view changes that severely degrade throughput.

\subsection{The Case for Learning}

Our motivating experiments above demonstrate that while the \emph{correctness} of DAG-based BFT protocols is never impacted by a configuration or environmental change, such changes can have large \emph{performance} impacts. We make two concrete observations. First, \textbf{no single configuration is universally optimal}: a configuration that maximizes performance under one environmental condition often performs poorly under another. In dynamic environments, relying on a fixed default configuration inevitably leads to substantial performance degradation. Second, \textbf{protocol parameters interact intricately} with one another and with the environment, forming complex dependencies that are difficult to model explicitly. If there is a simple heuristic or hand-crafted cost model that can correctly tune DAG-based BFT algorithms for a variety of different environmental conditions, we were unable to build it. Even if one were able to develop such a heuristic, it would be inherently limited to past experience, possibly failing to generalize to unseen or adversarial conditions.
Collectively, these observations point toward a clear direction: \textit{DAG-based BFT protocols must evolve toward an autonomous, learning-based framework to achieve robust performance across diverse and dynamic conditions.}
\section{Overview}\label{sec:overview}

\subsection{System Model}
In \sys, we consider a system consisting of $n = 3f + 1$ replicas and a set of clients, where up to $f$ replicas may exhibit Byzantine behavior. Each replica simultaneously serves two roles: a \textit{validator} and a \textit{learning agent}. The validator is responsible for data dissemination and block ordering, while the learning agent operates off the critical path to collect runtime observations, train learning models, and recommend configuration adaptations during execution.

A replica is considered correct if it follows the protocol specification; otherwise, it's considered faulty. We assume a strong Byzantine adversary that can arbitrarily coordinate all faulty replicas but cannot violate standard cryptographic assumptions. When a replica is faulty, it can behave arbitrarily in any of its roles: faulty validators may exhibit Byzantine behaviors, including equivocation, double voting, and selective message suppression; faulty learning agents may launch learning-specific attacks by reporting manipulated local observations, such as forged state features or rewards, to interfere with distributed model training.

\sys assumes a partially synchronous network model~\cite{dwork1988consensus}. Specifically, there exists an unknown Global Stabilization Time (GST), after which all messages between correct replicas are delivered within a bounded delay $\Delta$. Participants communicate with each other through reliable, authenticated, point-to-point channels. For two different roles on the same node, we assume their communication in-between is always synchronous.

\subsection{Design Overview}

\sys formulates the configuration tuning process as a reinforcement learning task with two key components:
(i) a reinforcement learning agent that guides the selection of parameter configurations according to the changing environment, and
(ii) a coordination protocol that orchestrates data collection in a distributed manner. \sys operates in epochs, each consisting of the execution of \(h\) slots (i.e., proposals in leader-based BFT protocols), during which the protocol configuration remains unchanged. Here, \(h\) is a predefined system hyperparameter that controls how frequently \sys adapts its configuration.

\begin{figure}[t]
    \centering
    \includegraphics[width= 0.48\textwidth]{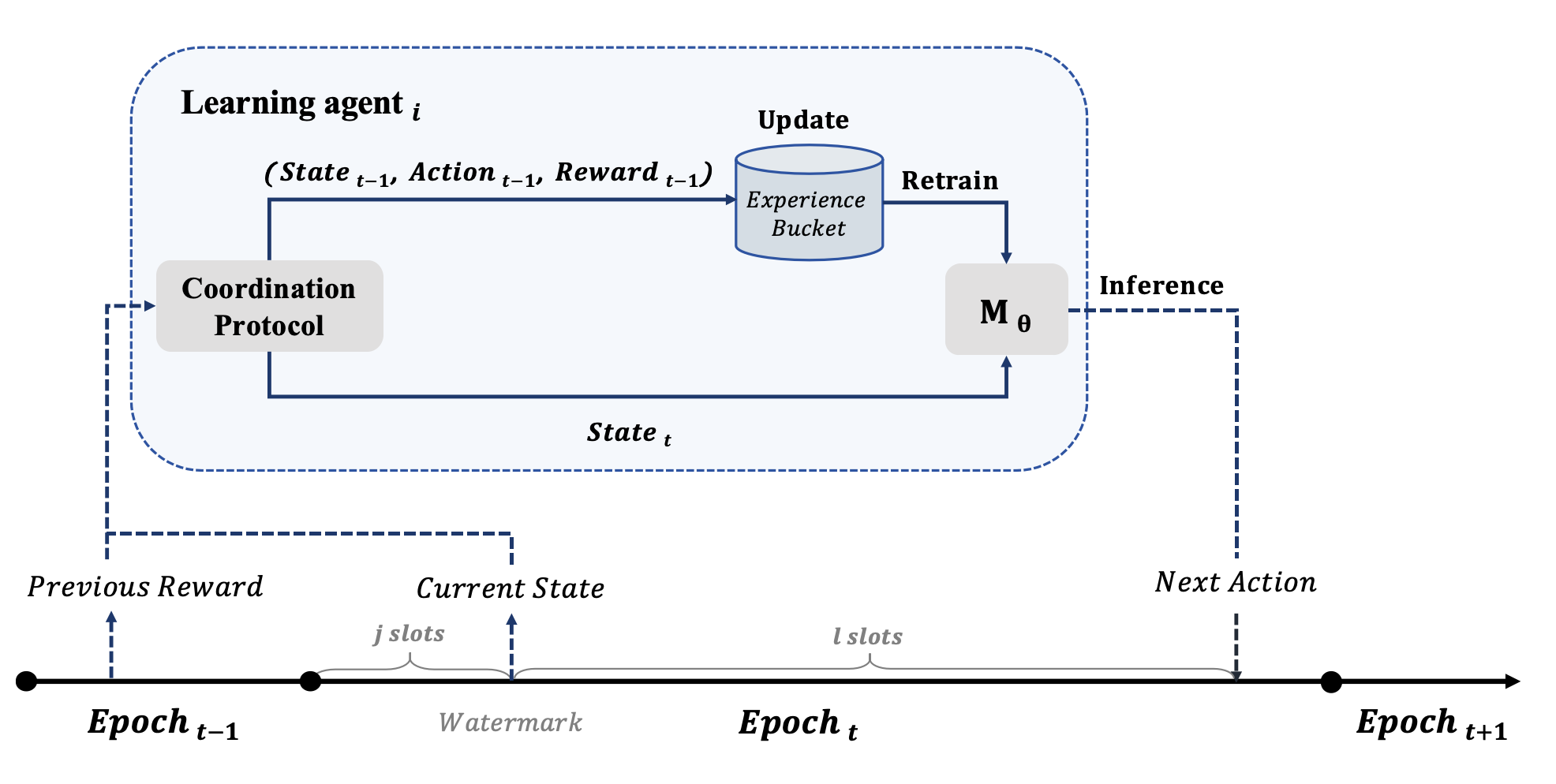}
    \caption{Overview of learning process on replica $i$.}
    \label{fig:workflow}
\end{figure}

\noindent \textbf{Learning agent.} Instead of relying on a centralized agent, \sys equips each validator with a local learning agent. Each agent formulates the configuration tuning problem as a contextual multi-armed bandit (CMAB) problem. In \sys, each agent periodically observes the current operating environment as the context, and selects one parameter configuration from a discrete set of candidate configurations (i.e., the arms). After the selected configuration takes effect, the agent observes the resulting system performance as the reward. Each observed \((\textit{context}, \textit{action}, \textit{reward})\) tuple constitutes a training data point that is used to refine the agent's policy over time. To be successful, \sys must balance the exploration of new, untested configurations with exploiting past experiences to maximize performance. We adopt the CMAB formulation rather than full reinforcement learning for two reasons. 
First, CMAB admits asymptotically optimal algorithms (e.g., Thompson Sampling~\cite{thompson_intro}) with well-studied regret bounds~\cite{thompson_bound}, providing strong theoretical guarantees on convergence. Second, CMAB algorithms are significantly more sample-efficient than full reinforcement learning, which is critical in a production system where each configuration trial incurs real performance cost. 
Details about the learning algorithms are provided in Section {\ref{sec:learning}}. 

Since \sys operates in a Byzantine environment where no centralized entity can be trusted, each validator only accepts configuration decisions from its co-located agent. The learning agents themselves collectively form a replicated state machine. Specifically, all agents are initialized from the same model state and random seed. For each epoch $t$, as detailed in Section~\ref{sec:coordination}, the underlying consensus protocol guarantees that all benign agents observe the same sequence of training inputs. With deterministic training, benign learning agents host the same model parameters for a given epoch. Since the learning process is replicated across all nodes, when different agents observe the same context at inference time, they independently derive the same configuration decision.

\noindent \textbf{Robust distributed data collection.} To prevent Byzantine nodes from polluting or equivocating training data, \sys introduces a robust coordination protocol.
After executing a predefined watermark of $j$ blocks in epoch $t$ (where $j < h$), each agent measures its local context and reward and broadcasts them to other replicas. By leveraging Autobahn's external consensus layer, the coordination protocol then establishes a globally consistent collection of reports, comprising inputs from at least a two-thirds quorum of agents. Each agent then applies an identical, deterministic filter to this agreed-upon aggregation to construct an identical training data point for subsequent learning. 
Details about this learning-coordination protocol are provided in Section \ref{sec:coordination}.

\noindent \textbf{\sys workflow. }Figure~\ref{fig:workflow} illustrates the workflow of \sys on replica $i$. Each replica in the system follows the same procedure. 

\noindent\textbf{Step 1: Online data collection.} After executing a predefined watermark of $j$ blocks in epoch $t$ (where $j < h$), each node $i$ measures its local context $s_{i, t}$ in the current epoch and reward $r_{i, t-1}$ observed during the previous epoch. Each agent then broadcasts its local observations to all other agents. By leveraging the learning coordination protocol, it yields a Byzantine-robust global context $S_{t}$ and global reward $R_{t-1}$ that all benign agents agree upon.

\noindent\textbf{Step 2: Model update.} Each agent adds the newly observed tuple $(S_{t-1}, A_{t-1}, R_{t-1})$ to its experience bucket, which stores all of the state-action-reward tuples observed at run-time. The agent then retrains the predictive model using the updated experience bucket.

\noindent\textbf{Step 3: Action selection.} Each agent independently infers the action for the current epoch using $M_\theta$ based on the current global context $S_{t}$. Since both $S_{t}$ and $M_\theta$ are consistent across all benign agents, the selected action $A_{t}$ is identical without requiring explicit coordination on action. After another $l~(j + l < h)$ slots, each validator switches to the newly selected parameter configuration, and the resulting reward will be measured in the upcoming epoch.

By repeating the above procedure in every epoch, each agent continuously refines $M_\theta$, progressively establishing a mapping from observed system conditions to optimal configurations and enabling fully adaptive parameter tuning at runtime.

\vspace{1em}
\section{Learning Algorithm}\label{sec:learning}
This section presents \sys's learning approach in detail. We first formalize the learning problem and explain how Thompson sampling solves it. We then outline the state and action space design, followed by the predictive model used by \sys.

\subsection{Problem Formulation}

\noindent \textbf{Contextual multi-armed bandits.} \sys formulates configuration tuning as a contextual multi-armed bandit (CMAB) problem, where each learning agent observes the current system state \(s_t\), selects a configuration \(a_t\) from a candidate action space, and observes the resulting performance \(r_t\) as reward. The objective of the agent is to maximize the cumulative reward over time. Depending on the application that leverages BFT consensus, the reward function can be defined using different performance metrics, such as throughput, latency, or a weighted combination of both. 

CMABs assume that epochs are independent from each other, and the optimal action depends only on the current state. \sys satisfies this assumption, as the current choice of parameter configuration does not affect the pattern of workloads, network, or faults in future epochs. Under this formulation, each epoch is modeled as an independent decision round conditioned on the observed state.

\noindent \textbf{Thompson sampling.} 
\sys adopts Thompson Sampling to solve contextual multi-armed bandit problems for its simplicity: at the beginning of each decision epoch, an agent retrains a predictive model using the experiences accumulated in its experience bucket, and then selects the best action predicted by the model. Rather than relying on a model that best fits the observed experience, the sample model parameters are proportional to their likelihood given the training data. More formally, we can define maximum likelihood estimation as finding the model parameters $\theta$ that maximize likelihood given experience $E$: $\argmax_\theta P(\theta \mid E)$ (assuming a uniform prior). Instead of maximizing likelihood, Thompson sampling simply samples from the distribution $P(\theta \mid E)$. This means that regions of the parameter space strongly supported by the observed data are assigned higher posterior probability and therefore more likely to be sampled. In contrast, regions with less empirical support receive lower posterior probability and are sampled less frequently.

\subsection{State, Action and Reward}
Guided by the observations in Section~\ref{sec:motivation}, we identify the runtime factors that most significantly affect Autobahn's performance and use them to construct the state space. We define the action space based on the tunable protocol parameters exposed by Autobahn.

\subsubsection{State Space}
To provide the predictive model with sufficient context, the state space should capture the operating environment of \sys, such as workload characteristics and network conditions. Directly measuring low-level network conditions, however, is challenging in large-scale Byzantine environments due to potential adversarial behaviors. Instead, \sys leverages signals naturally embedded in Autobahn’s local lane structure and consensus execution. Our state space consists of two components.

\noindent\underline{\textit{State 1: Lane Growth Rate.}} We characterize Autobahn’s data dissemination layer as a lane-based structure rather than a strict DAG structure. At time \(\tau\), node \(i\) maintains a local view of the committed lane lengths:

\begin{equation}
\mathbf{L}_\tau^{(i)} =[l_{1,\tau}^{(i)}, l_{2,\tau}^{(i)}, \dots, l_{n,\tau}^{(i)}]
\end{equation}

where \(l_{j,\tau}^{(i)}\) denotes the number of committed blocks in validator \(j\)'s lane as observed by node \(i\).

To capture the dissemination progress, \sys computes the lane growth-rate vector at time $\tau$ over a sliding observation window of duration \(\Delta \tau\):

\begin{equation}
\mathbf{R}_\tau^{(i)} =
\frac{
\mathbf{L}_\tau^{(i)} -
\mathbf{L}_{\tau-\Delta \tau}^{(i)}
}{\Delta \tau}
\end{equation}

where each entry \(r_{j,\tau}^{(i)}\) represents the average growth rate of validator \(j\)'s lane during the observation window.

Intuitively, lane growth rate is jointly determined by (i) workload assigned to each node, (ii) the latency required to collect the first $f+1$ votes for a newly proposed header, and (iii) node processing capability that affects header production rate. Replicas with higher transaction arrival rates, lower communication latency to neighboring replicas, or better processing capability tend to extend their lanes more rapidly, while the remaining nodes might fall behind. Although the growth rate of an individual lane only reflects the operating conditions of a particular replica, the lane growth-rate vector $\mathbf{R}_\tau^{(i)}$ captures the growth rate of all lanes, providing an overall view of workload distribution, network conditions, and replica heterogeneity across the system.

\noindent\underline{\textit{State 2: Fast-path ratio.}}
We define the fast-path ratio as the proportion of slots finalized through the fast path of Autobahn's external consensus within the observation window. This metric reflects the network condition of the system: a high fast-path ratio indicates that the communication delays among replicas are generally bounded by the configured fast-path timeout, allowing the required $3f+1$ votes to be collected before the timeout expires. Conversely, a low fast-path ratio implies slower communication for some nodes, causing requests to fall back to the slow path more frequently. The feature also captures the impact of faults, such as omission failures, leader equivocation, and network partitions, which can all reduce fast-path success and thereby degrade performance.

Together, these two state features provide an expressive representation of \sys's operating environment. Importantly, these signals are \emph{naturally observable} from the standard execution of the consensus protocol without any additional instrumentation, explicit network measurement, or extra communication overhead. This makes the state construction lightweight and efficient when deployed in high-performance BFT systems.

\subsubsection{Action Space}
We define the action space by selecting representative and influential tunable knobs divided into two categories according to the protocol structure: data dissemination and consensus. Each action dimension is discretized into a finite set of candidate values, resulting in a total of 72 possible actions.

\noindent\textbf{Data dissemination.} This category controls how transactions are propagated and how lanes are built.

\noindent\underline{\textit{Actions 1\&2: Batch size and Max batch delay.}}
These two parameters jointly determine the transaction batching behavior before dissemination. Batch size determines how many transactions are aggregated into a batch, while max batch delay bounds the waiting time before a partially filled batch is sealed and broadcast. Together, they balance responsiveness against communication and cryptographic overhead. Small values lead to frequent dissemination of partially filled batches, increasing network and signature verification overhead. Conversely, overly large values increase transaction queuing delay before dissemination, increasing end-to-end latency.

\noindent\underline{\textit{Actions 3\&4: Header size and Max header delay.}}  
These two parameters primarily affect how fast the lane grows and how frequently consensus is triggered. Header size determines how many batches are aggregated into a header, while max header delay limits how long it can be postponed before a header is sealed.  Aggressive configurations generate headers rapidly, causing frequent consensus executions with limited useful payload per proposal and poor amortization of consensus overhead. In contrast, conservative configurations slow lane expansion and delay proposal generation, increasing the time transactions remain pending before execution. Therefore, these parameters govern the trade-off between consensus efficiency and proposal responsiveness.

Since both batch size and max batch delay play similar roles in determining when a batch is sealed, jointly tuning parameters concurrently would introduce an additional action dimension while providing limited additional control over the batching behavior. We therefore fix the max batch delay to a sufficiently large value (e.g., $4000$ ms), making batch formation primarily driven by batch size. Similarly, we fix the max header delay to $4000$ ms and focus on dynamically tuning the header size instead.

\noindent\textbf{Consensus.}
This category comprises parameters for optimizing the performance of the linear BFT consensus layer. 

\noindent\underline{\textit{Action 5: Cut condition.}} This parameter defines the number of new tips required to propose a new lane cut. A more aggressive cut condition may increase proposal frequency and responsiveness, but is at the cost of increasing the overhead amortized to transactions. In contrast, a conservative cut condition may commit more transactions each time, but is at the risk of excessive waiting time, especially under skewed DAG growth.

\noindent\underline{\textit{Action 6: Fast-path timeout.}} This parameter determines the maximum waiting time for collecting $3f+1$ \texttt{prepare} votes in the fast path. If the timeout expires before enough votes are received, the protocol falls back to the expensive slow path. Setting this value to zero effectively disables the fast path. Its optimal configuration heavily depends on communication latency and network stability.

\noindent\underline{\textit{Action 7: Number of
parallel proposals.}} This parameter controls the number of consensus slots allowed to be processed concurrently. 
Normally, increasing the degree of parallelism allows multiple slots to make progress concurrently, reducing latency and improving resource utilization. However, since proposals are ultimately committed in sequence, excessive pipelining can increase contention for CPU, network, and memory resources, causing the benefits of additional parallelism to decrease or even backfire.

\subsubsection{Reward Function}
Modern DAG-based consensus protocols have substantially improved throughput via highly parallelized data dissemination. As a result, system performance is often no longer primarily limited by throughput, but by end-to-end latency, which becomes the dominant bottleneck for production use cases (e.g., latency-sensitive decentralized finance applications). Therefore, we define a reward function to optimize end-to-end latency as follows:
\begin{equation}
    r = \frac{1}{latency}
\end{equation}

where latency is measured in seconds. Our objective is to maximize the reward $r$, i.e., to minimize end-to-end latency.

\vspace{0.5em}
\subsection{Predictive model}
\vspace{0.2em}

\sys chooses random forests~\cite{breiman2001random} as the predictive model due to their good performance on data sets of moderate sizes and fast inference. Each learning agent hosts a predictive model that follows the \textit{value-based} RL approach: given the featurized state, predict the performance (i.e., reward) of each parameter configuration (i.e., action). If there is a tie on the best-predicted performance, \sys breaks the tie randomly to avoid local maxima. 

Integrating a predictive model with Thompson sampling requires the ability to sample model parameters from $P(\theta | E)$, the distribution of model parameters given the current experience. The simplest technique (which has been shown to work well in practice~\cite{thompson_bootstrap}) is to train the model as usual, but only on a bootstrap~\cite{bootstrapping} of the training data. More specifically, the predictive model is trained on a bootstrap dataset, constructed by randomly sampling $|E|$ experiences with replacement from the experience set. This mechanism naturally provides the exploration-exploitation trade-off required by contextual multi-armed bandits. During the early training stage, the experience set is relatively small and sparse, causing the bootstrap datasets to differ significantly from one another. As a result, the variance of prediction is large, and the agent may occasionally overestimate the reward of underexplored actions. Such variability encourages exploration of diverse configurations. As training progresses and more experiences are accumulated, the uncertainty of reward estimation gradually decreases. Consequently, predictions produced by different bootstrap models become increasingly consistent, causing the agent to repeatedly select configurations with the highest estimated rewards and thereby favor exploitation.
\vspace{0.5em}
\section{Learning Coordination Protocol}\label{sec:coordination}
\vspace{0.5em}

Since \sys collects observations from all participants to guide online learning, Byzantine nodes may manipulate their reported observations to misguide the learning process (e.g., over-reporting or under-reporting local measurements). To address this challenge, \sys adopts a robust coordination protocol, aimed at constructing a consistent and accurate global view of observations despite the presence of up to \(f\) faulty nodes. We further decompose the observations into two categories: local and global. Local ones are observed independently by each node based on its view of the system. In contrast, global ones represent the consistent observations the whole system agrees upon, constructed by aggregating local observations across nodes through the coordination protocol. 

As illustrated in Algorithm~\ref{alg:coordination}, \sys implements a lightweight coordination protocol that reuses Autobahn’s external BFT consensus layer without introducing additional consensus rounds. The only additional communication occurs when each node broadcasts a \(LocalReport\) once per epoch. At epoch \(t\), when replica \(i\) reaches the observation threshold \(j\), it computes its local runtime measurements \(LocalState_i\) and \(LocalReward_i\), and broadcasts \(LocalReport_t\) to all other replicas. The node simultaneously starts a local collection timer \(T_i^t\) (Lines 2-6).

Each node maintains a local report buffer \(Reports_i^t\) containing all received \(LocalReport_t\)s. Once a node collects $3f+1$ reports for epoch $t$, or the local collection timer \(T_i^t\) expires after receiving at least $2f+1$ reports, it locally constructs a \(ReportSet_t\) by assembling the collected reports, without any need to forward reports to a designated aggregator (Lines 7-11). 

Any honest leader can immediately embed its locally available  \(ReportSet_t\) into a proposal upon obtaining a proposal opportunity under the round-robin scheme. The \(ReportSet_t\) is then piggybacked onto the next proposal in Autobahn's consensus layer. Once the proposal is committed, all correct replicas obtain the same certified aggregation of reports, denoted as \(CertifiedReport_t\)  (Lines 12--19). Once the \(CertifiedReport_t\) is obtained, each node deterministically extracts global states and rewards from the set of reports. Specifically, \sys applies different aggregation strategies to different observations for robustness. For lane growth rates, \sys uses max aggregation to reconstruct the most up-to-date lane progression. For $fast~path~ratios$ and rewards, \sys adopts median aggregation to tolerate Byzantine outliers. The aggregated observations constitute a globally consistent  \(GlobalReport_t\), including \(GlobalState_t\) and \(GlobalReward_t\) for epoch $t$ (Lines 20--28).

\begin{algorithm}[t]
\caption{Learning Coordination Protocol}
\label{alg:coordination}
\begin{algorithmic}[1]
\small
\State \textbf{State:} epoch $t$, node id $i$

\Statex
\Statex {\color{blue}\textbf{$\triangleright$ Phase 1: Local Observation}}

\Upon{executed slots at node \(i\) reach observation threshold \(j\)}
    \State \(LocalState_i \gets \textsc{ComputeState}(local\ execution\ logs)\)
    \State \(LocalReward_i \gets \textsc{ComputeReward}(local\ execution\ logs)\)
    \State broadcast $LocalReport(t, i, LocalState_i, LocalReward_i)$
    \State start local timer \(T_i\)
\EndUpon


\Statex
\Statex {\color{blue}\textbf{$\triangleright$ Phase 2: Local Report Collection}}
\Statex \textsc{// Each node independently assembles reports.}

\Upon{receive
\(LocalReport(t, j, LocalState_j, LocalReward_j)\)}
    \If{\(t\) matches current epoch}
        \State store report into buffer \(Reports_i^t\)
    \EndIf
\EndUpon

\Upon{$|Reports_i^t| \ge 3f+1$ \textbf{or} ($T_i$ expires {and} $|Reports_i^t| \ge 2f+1$)}
    \State \(ReportSet_t \gets\) \textsc{AssembleReports}\((Reports_i^t)\)
\EndUpon

\Statex
\Statex {\color{blue}\textbf{$\triangleright$ Phase 3: Consensus Piggybacking}}
\Statex \textsc{// Disseminate and persist Aggregated Report}
\Upon{node \(i\) becomes proposer in Autobahn consensus}
    \If{\(ReportSet_t\) exists and is not yet committed}
        \State Embed \(ReportSet_t\) into proposal \(B\)
        \State Execute existing Autobahn linear consensus
    \EndIf
\EndUpon

\Upon{proposal \(B\) committed}
    \If{\(B\) contains \(ReportSet_t\)}
        \State \(CertifiedReport_t \gets B.ReportSet_t\)
        \State Garbage collect local \(ReportSet_t\)
    \EndIf
\EndUpon

\Statex
\Statex {\color{blue}\textbf{$\triangleright$ Phase 4: Robust Extraction}}
\Statex \textsc{// Feature-specific aggregation}
\Upon{$CertifiedReport_t$ obtained}
    \State Extract $\{LocalReward_i\}$, $\{r_j^{(i)}\}$, and $\{fast\_path\_ratio_i\}$
    \State $GlobalReward_t \gets \textsc{Median}(\{LocalReward_i\})$

    \For{each validator lane \(j\)}
        \State $ \hat{r}_{j} \gets \max \left( r_{j}^{(1)},  r_{j}^{(2)}, \dots, r_{j}^{(n)} \right)$
    \EndFor

    \State $\hat{\mathbf{R}} = [\hat{r}_{1}, \hat{r}_{2}, \dots, \hat{r}_{n}]$
    
    \State \(GlobalFastPathRatio_t \gets \textsc{Median}(\{fast\_path\_ratio_i\})\)
    
    \State \(GlobalState_t = \{ \hat{\mathbf{R}}_t, GlobalFastPathRatio_t \} \)

    \State \(GlobalReport_t\) $\gets$ \((GlobalState_t, GlobalReward_t)\)
    
    \State
    Export \((GlobalReport_t\) to the learning agent
    
\EndUpon
\end{algorithmic}
\vspace{1.0em}
\end{algorithm}

\vspace{0.5em}
\subsection{Proof Sketch}

\subsubsection{Liveness}
There exists an epoch $t$, at which every honest learning agent is guaranteed to invoke its learning process.

\noindent \textbf{Step 1: Report Collection Progress.}
Under the partial synchrony model, there exists a GST after which all messages sent by honest replicas are delivered within a bounded delay $\Delta$. Each honest replica broadcasts a \(LocalReport\) once per epoch upon reaching the block execution threshold $j$. A replica constructs an \(ReportSet\) upon either of the following two conditions: (i) it receives $3f+1$ local reports; or (ii) timer $T_i^t$ expires after a duration exceeding $\Delta$. In case (ii), the partial synchrony assumption guarantees that all messages from honest replicas have been delivered, ensuring that at least $2f+1$ honest reports are included in the received set. Therefore, every honest replica eventually constructs a valid $ ReportSet_t$.

\noindent\textbf{Step 2: Certification of Aggregated Reports.}
The epoch $t$ becomes learnable only after the \(ReportSet_t\) is certified via the consensus as the \(CertifiedReport_t\). The main challenge is that Byzantine leaders may intentionally omit \(ReportSet_t\) from their proposals. Since Autobahn’s linear BFT consensus employs round-robin leader rotation and there are at most \(f\) Byzantine replicas, within at most \(f+1\) consecutive views, there eventually exists an honest leader. Honest replicas hold all received reports locally until certification and never discard a valid \(ReportSet_t\) due to failed proposals, view changes, or leader replacement. Consequently, once an honest replica becomes the leader, it eventually includes \(ReportSet_t\) into a proposal block. By the liveness property of the underlying linear BFT consensus under partial synchrony, the proposal carrying a \(ReportSet\) will eventually be committed, thereby producing a certified coordination state \(CertifiedReport_t\).

\noindent\textbf{Step 3: Learning Invocation.}
Since $CertifiedReport_t$ is derived from the committed block, the liveness of Autobahn's underlying BFT consensus guarantees that every honest replica eventually receives and processes the same $CertifiedReport_t$. As a result, every honest learning agent eventually derives the same $GlobalReport_t$ and invokes the learning process at the same epoch $t$. 

\subsubsection{Safety.}

Since each $ReportSet_t$ is embedded in the consensus messages of Autobahn’s linear BFT protocol, the coordination protocol inherits the same safety guarantees. Consequently, all honest replicas derive the same $CertifiedReport_t$ for epoch $t$. Given the same certified input, deterministic aggregation produces identical $GlobalState_t$ and $GlobalReward_t$ values at every honest replica.


\subsubsection{Robustness.}
We prove here that the aggregated training data obtained by each learning agent is authentic and accurate despite the polluted reports from up to $f$ Byzantine nodes. First, we show that the max-based aggregation mechanism used to reconstruct lane structure correctly captures the newest progress achieved by honest replicas (Lemma~1). Second, we show that the aggregation mechanism used for other metrics ($fast~path~ratio$ and reward) is robust to Byzantine behaviors: even when up to $f$ nodes report arbitrary values, median-based aggregation ensures that the resulting fast path ratio and reward remain within the range of honest inputs (Lemma~2). Together, these results guarantee that both states and reward signals used for learning accurately and honestly reflect the system condition, despite Byzantine interference.

\noindent \textbf{Lemma 1.} Under max aggregation, the globally aggregated lane growth-rate is guaranteed to accurately reflect the true lane progress.

\noindent\textbf{Proof.}
Lane progression in Autobahn is cryptographically verifiable since every committed block must carry a valid Proof of Availability (PoA). Therefore, Byzantine replicas cannot over-report non-existent lane progress because they are unable to forge valid certificates or signatures. A malicious replica may only under-report its locally observed lane progress by omitting valid committed blocks.

The globally aggregated lane growth rate is defined as:

\begin{equation}
\hat{r}_{j}
=
\max_{r \in ReportSet_t}
r[j].
\end{equation}

Since Autobahn enforces in-order voting and requires at least \(f+1\) votes before extending a lane (i.e., proposing a new CAR), at least one honest replica must possess the latest valid committed history for every lane. Moreover, under the partial synchrony assumption, there exists a bounded message delay \(\Delta\) after GST such that all messages sent by honest replicas are delivered within \(\Delta\). During report collection, each replica starts a local collection timer after broadcasting its \(LocalReport\), which is configured to exceed the bounded network delay after GST. Therefore, by waiting for either all \(3f+1\) reports or timeout expiration, every honest replica eventually collects all reports sent by honest nodes. In other words, there exists at least one honest report \(r^\star \in Reports^t\) such that:

\begin{equation}
r^\star[j]
=
r_j^{true},
\end{equation}

where \(r_j^{true}\) denotes the true lane growth rate of validator lane \(j\).

Since Byzantine replicas can only under-report but cannot fabricate larger valid lane progress values, no report can exceed the true committed growth rate. Consequently,

\begin{equation}
\hat{r}_{j}
=
\max_{r \in Reports^t}
r[j]
=
r_j^{true}.
\end{equation}

Therefore, the lane growth-rate vector produced by max aggregation strategy accurately captures the latest lane progress despite the presence of up to $f$ Byzantine replicas.

\noindent \textbf{Lemma 2.} \label{lem:aggregation}
Under median aggregation, if all honest $fast~path~ratio$/ reward values form a range \([range_l, range_h]\), then the global taken aggregated value is also guaranteed to always fall into this range.

\noindent\textbf{Proof.}
Byzantine replicas arbitrarily report extremely large or small values on $fast~path~ratio$ or reward values. Let $R$ denote the set of reports in $ReportSet_t$, where $|R| \geq 2f+1$, and at most $f$ values may be arbitrarily manipulated by Byzantine replicas. Let $\mathrm{median}(R)$ denote the lower median of $R$, i.e., the $\lceil |R|/2 \rceil$-th smallest element. We assume all honest values lie within the interval $[range_l, range_h]$, and that honest replicas always report valid (non-null, non-zero) values.
Suppose the median value $r_m$ satisfies $r_m > range_h$. Then at least $\lceil |R|/2 \rceil$ elements in $R$ must be strictly greater than $r_m$. However, since $|R| \ge 2f+1$ and follows that $\lceil |R|/2 \rceil > f$, at most $f$ Byzantine replicas alone to contribute $f+1$ such values. This implies that at least one honest value must exceed  $[range_h]$, contradicting the assumption that all honest values lie within $[range_l, range_h]$. A symmetric argument shows that $r_m \ge range_l$. Therefore, the median is guaranteed to lie within the range of honest observations.

\section{Evaluation} \label{sec:evaluation}

\begin{figure*}[!t]
\begin{minipage}{0.33\textwidth}
	\centering
	\includegraphics[width=\columnwidth]{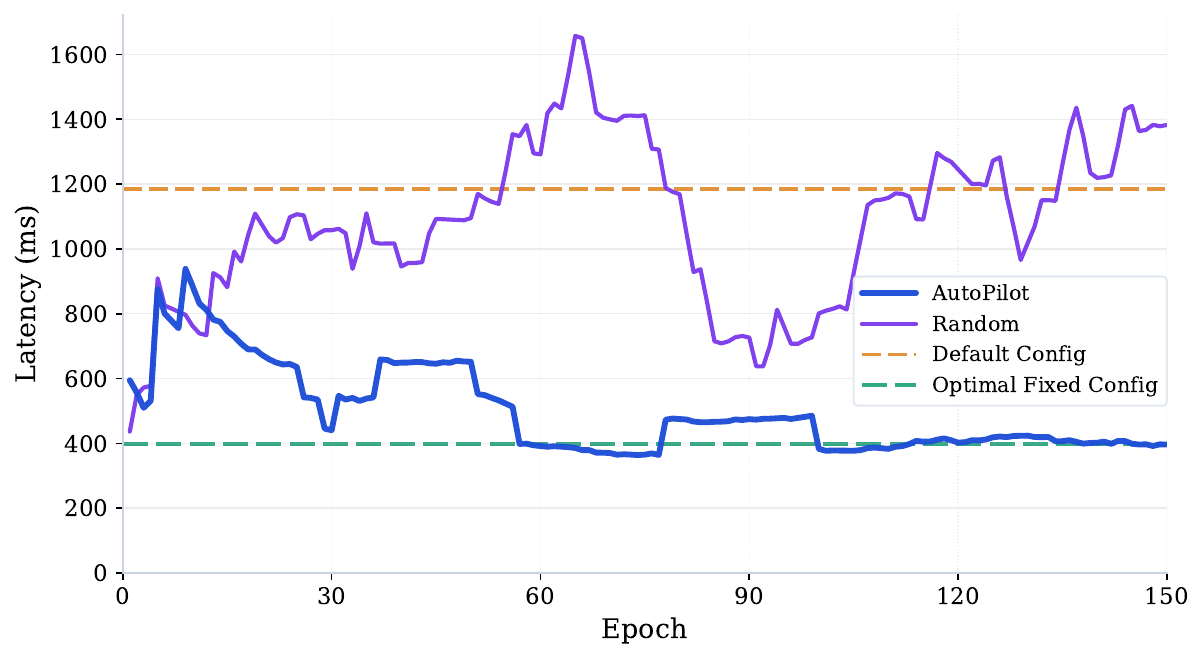}
    \vspace{-2.2em}
    \captionsetup{format=plain, font=scriptsize}
    \caption*{\normalfont \,\,\,\,\, (a) Scenario $S_1$}
\end{minipage}
\begin{minipage}{0.33\textwidth}
	\centering
	\includegraphics[width=\columnwidth]{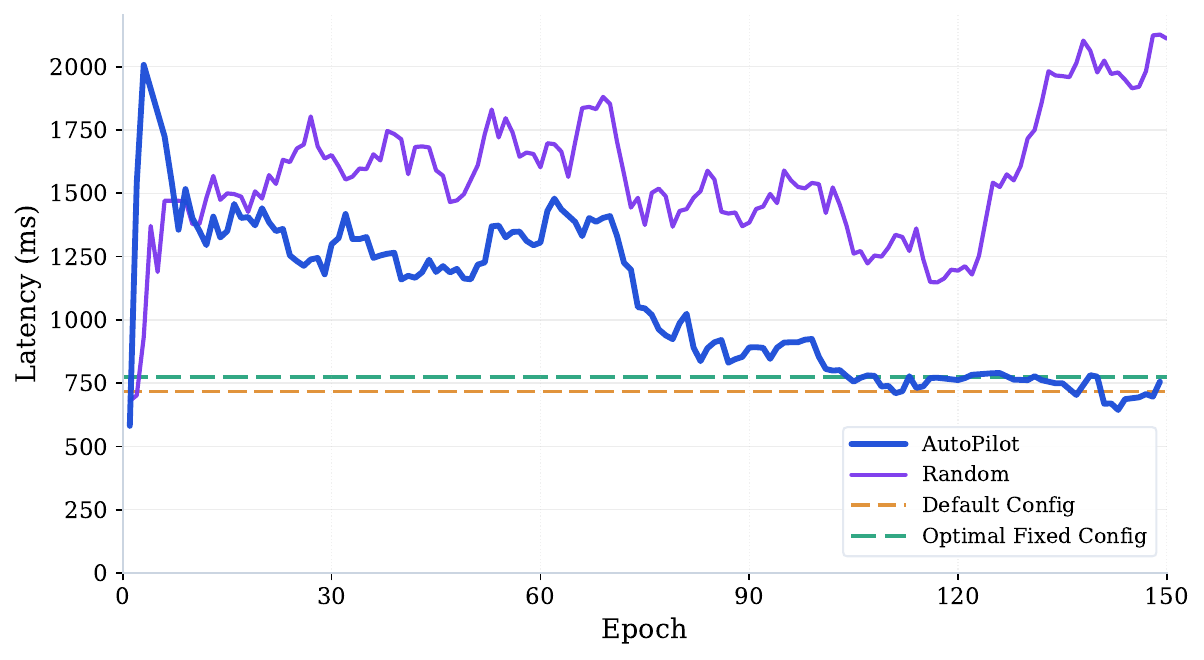}
    \vspace{-2.2em}
    \captionsetup{format=plain, font=scriptsize}
    \caption*{\normalfont \,\,\,\,\, (b) Scenario $S_2$}
\end{minipage}
\begin{minipage}{0.33\textwidth}
	\centering
	\includegraphics[width=\columnwidth]{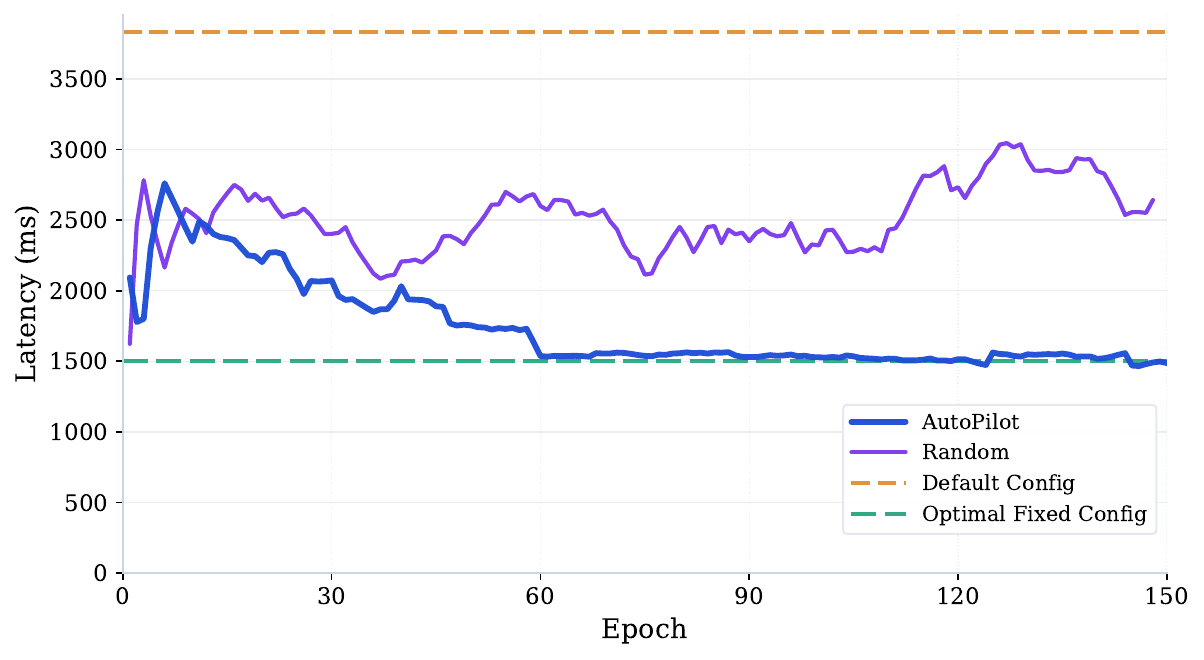}
    \vspace{-2.2em}
    \captionsetup{format=plain, font=scriptsize}
    \caption*{\normalfont \,\,\,\,\, (c) Scenario $S_3$}
\end{minipage}

\captionsetup{labelfont={color=black, bf}}
\caption{Convergence of \sys in three static scenarios. Each plot shows the performance of \sys, the default configuration, random policy and the optimal fixed configuration for that scenario. While the optimal parameter configuration differs for each scenario, \sys always reaches near-optimal performance.}
\label{fig:static}
\end{figure*}

Our evaluation aims to answer the following questions:

\noindent (1) Can \sys converge to the optimal configuration under a static scenario without pre-training? (Section \ref{subsec:static})

\noindent (2) How effectively does \sys adapt to dynamically changing environment, compared with the optimal static configuration and default configuration? (Section~\ref{subsec:dynamic})

\noindent (3) How robust is \sys against adversarial data pollution, and how much does its robust learning coordination protocol improve performance? (Section~\ref{subsec:robustness})

\noindent (4) What overhead does \sys introduce? (Section \ref{subsec:overhead} )

In the rest of this section, we present the experimental setup
and then answer each of the above questions.

\subsection{Experimental Setup}

\noindent\textbf{Implementation}.
We have implemented a prototype of \sys with Rust and Python, starting from the open source implementation of Autobahn\cite{autobahn-codebase}. It uses Tokio TCP\cite{tcp} for networking, RocksDB \cite{rocksdb} for persistent storage, and ed25519-dalek\cite{ed25519-dalek} signatures for authentication. On each node, \sys runs three independent processes: a Rust-based consensus protocol, a Python metrics collector that periodically parses execution logs to compute runtime states and rewards, and a Python controller that hosts the online learning agent. Each consensus process communicates with its co-located Python processes through Unix domain socket channels. Our implementation is publicly available\footnote{\url{https://github.com/ccclr/AutoPilot}}.

\noindent\textbf{Testbed}. 
Our testbed consists of Google Cloud Platform (GCP) \texttt{t2d-standard-4} instances equipped with 4 vCPUs, 16\,GB of memory, and 10\,Gbps network bandwidth. Our experiments are conducted with $n = 4$ replicas, where one replica is deployed in \texttt{asia-east2-a}, one replica is deployed in \texttt{us-central1-c}, and the remaining two replicas are deployed in \texttt{us-central1-f}. Each replica is co-located with a client process that continuously submits no-op transactions consisting of 512 random bytes. Unless otherwise stated, clients generate transactions at a constant arrival rate throughout the experiments. All reported latency measurements correspond to end-to-end commit latency observed by clients.

\begin{figure*}[!t]
    \centering
    \includegraphics[width= 1.0\textwidth]{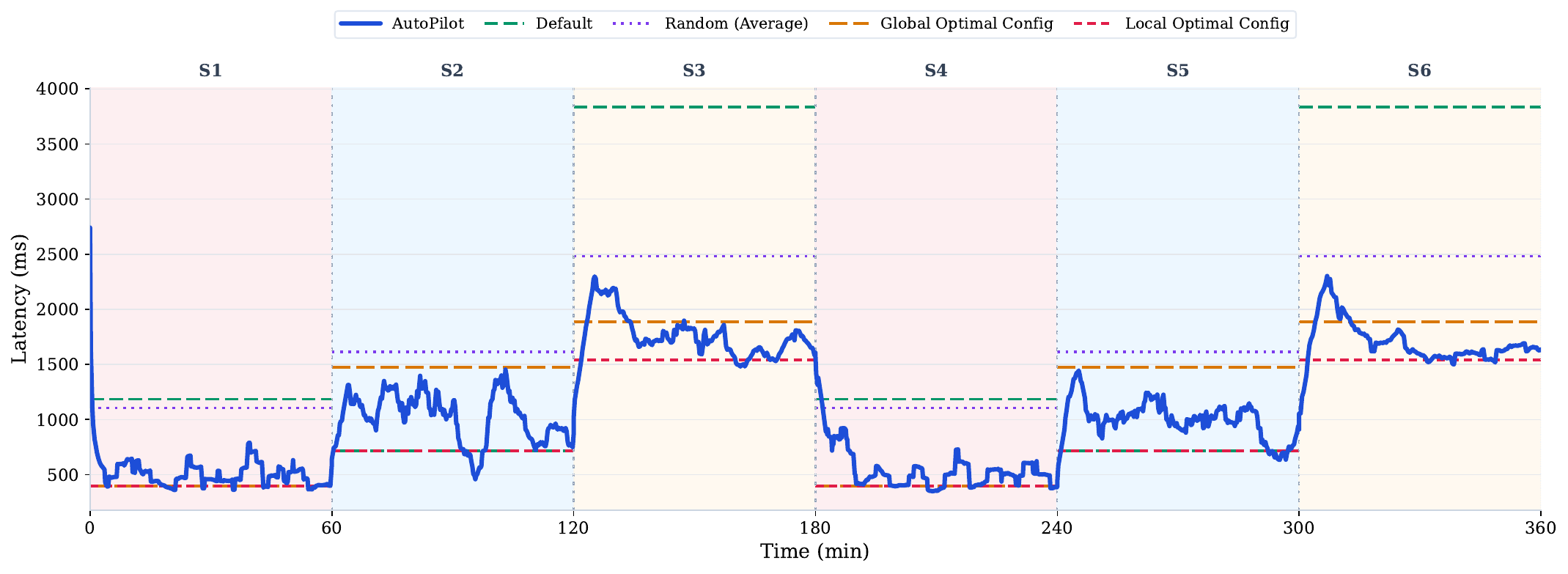}
    \caption{Adaptivity of \sys under changing conditions. The vertical dashed lines indicate when environments change.}
    \label{fig:dynamic}
\end{figure*}

\begin{table}[!t]
\centering
\caption{Static scenarios}
\label{tab: static scenario}

\scriptsize
\setlength{\tabcolsep}{10pt}

\begin{tabular}{ccc}
\toprule
\textbf{Scenario} & \textbf{Workload} & \textbf{Message Delay} \\
\midrule
\textbf{S1 (Normal case)} & Uniform & None \\
\textbf{S2 (Imbalanced workload)} & One node at \(1/100\) tx rate & None \\
\textbf{S3 (Degraded network)} & Uniform & 550\,ms / 750\,ms \\
\bottomrule
\end{tabular}
\vspace{2em}
\end{table}

\subsection{Convergence under Static Conditions}\label{subsec:static}
Our first set of experiments evaluates whether \sys can automatically converge to the optimal protocol configuration under different static scenarios, without relying on any prior training data. We pick three representative scenarios as illustrated in Table \ref{tab: static scenario}, each of which is associated with a different optimal configuration. In Scenario $S_1$, all nodes receive transactions at the same arrival rate and no additional consensus message delay is injected. This setting represents a stable deployment environment. In Scenario $S_2$, one node receives transactions at only \(1/100\) of the arrival rate of the remaining nodes, while no additional network delay is injected. This setting simulates workload imbalance and hotspot-style traffic distributions commonly observed in realistic deployments. In Scenario $S_3$, all nodes receive transactions at the same arrival rate, while additional consensus message delays are injected into two nodes (550\,ms and 750\,ms, respectively). This setting represents degraded network conditions arising from either network fluctuation or intentional delay by Byzantine nodes.
For each scenario, we run \sys for 160 epochs on a WAN and compare it against three baselines: (1) the optimal fixed configuration obtained via offline search, (2) a policy that randomly selects configurations, and (3) the default configuration used by Autobahn. 

Figure~\ref{fig:static} illustrates the convergence behavior of \sys under different scenarios by plotting the latency throughout online training. Each plotted point represents the average latency over the most recent 20 epochs.
Despite starting without any pre-training or prior knowledge, \sys consistently converges towards optimal parameter configurations across all scenarios. 
In our evaluation, convergence is defined as the point where the moving-average latency over the most recent 20 epochs remains consistently close to the best static configuration for multiple consecutive epochs. Under this definition, \sys converges substantially earlier than exhaustive exploration across all scenarios. In $S_1$, \sys reaches stable near-optimal performance after approximately 60 epochs (about 12 minutes). In $S_2$, convergence occurs after approximately 90 epochs and requires about 28 minutes due to the larger execution latency introduced by severe workload skew. In $S_3$, \sys converges after approximately 60 epochs, yet the convergence time remains around 25 minutes because injected network delays increase the duration of each learning epoch to roughly 14 seconds. 

In the early stages, the bootstrap datasets constructed from limited experience differ substantially from one another, producing high predictive variance across configurations. This uncertainty drives exploration, but once a configuration demonstrates consistently high reward, its estimated reward remains high in the bootstrap samples that include it, causing Thompson Sampling to select it more frequently than unknown configurations. As a result, \sys gradually concentrates on promising configurations rather than distributing equal budget across the entire action space as exhaustive searching or random selecting would. As experience accumulates, bootstrap variance decreases and predictions stabilize, shifting the agent toward exploitation of the best-identified configuration. Consequently, \sys reaches near-optimal performance after evaluating only a fraction of the action space. Consequently, compared with random configuration selection, \sys significantly reduces the average latency throughout execution.

\subsection{Adaptivity under Changing Conditions} \label{subsec:dynamic}

We next evaluate \sys under continuously changing deployment conditions. To emulate a changing environment, we run Scenarios~$S_1$--$S_3$ sequentially for 60 minutes each in a round-robin manner, and repeat the sequence after Scenario $S_3$. We compare \sys against four baselines: (1) the locally optimal fixed configuration for each individual scenario; (2) the globally optimal fixed configuration, selected as the configuration with the highest average performance across all three scenarios; (3) a random exploration policy; and (4) the default configuration used by Autobahn.

Figure~\ref{fig:dynamic} shows the latency averaged over a sliding window of the most recent 20 epochs throughout the 6-hour experiment. As deployment conditions change, \sys consistently adapts its configuration and converges toward the best-performing policy for the current environment. Overall, \sys achieves a 13.2\% reduction in average end-to-end latency compared with the fixed global optimal configuration, a 49.8\% reduction compared with Autobahn’s default configuration, and a 73.3\% reduction compared with the random exploration policy.


These results demonstrate two important advantages of \sys. First, \sys not only outperforms both globally fixed configurations and Autobahn’s default policy, but also eliminates the need for expensive offline data collection and pre-training prior to deployment. Second, \sys demonstrates strong adaptability across fundamentally different operating conditions, including workload heterogeneity (Scenario $S_2$) and faulty network environments with intentional delay injection (Scenario $S_3$). In particular, under highly heterogeneous workloads, \sys achieves up to 44.9\% lower latency compared to the best global fixed configuration.

Conducting an exhaustive offline grid search to identify the optimal configuration is expensive since it requires evaluating a large number of candidate configurations under each deployment condition. In the worst case, the search procedure must explore nearly the entire action space before identifying the optimal configuration, resulting in a performance similar to random exploration. Compared with \sys, the random exploration policy incurs significantly higher average latency across all scenarios: 129.4\% higher in $S_1$, 57.3\% higher in $S_2$, and 50.3\% higher in $S_3$.

To balance exploration and exploitation, we limit the experience set to the most recent 180 observations. This provides sufficient data for convergence under static scenarios while reducing the impact of outdated observations, enabling \sys to adapt more effectively to changing operating conditions.

\begin{figure}
\begin{minipage}{0.4\textwidth}
	\centering
	\includegraphics[width=\columnwidth]{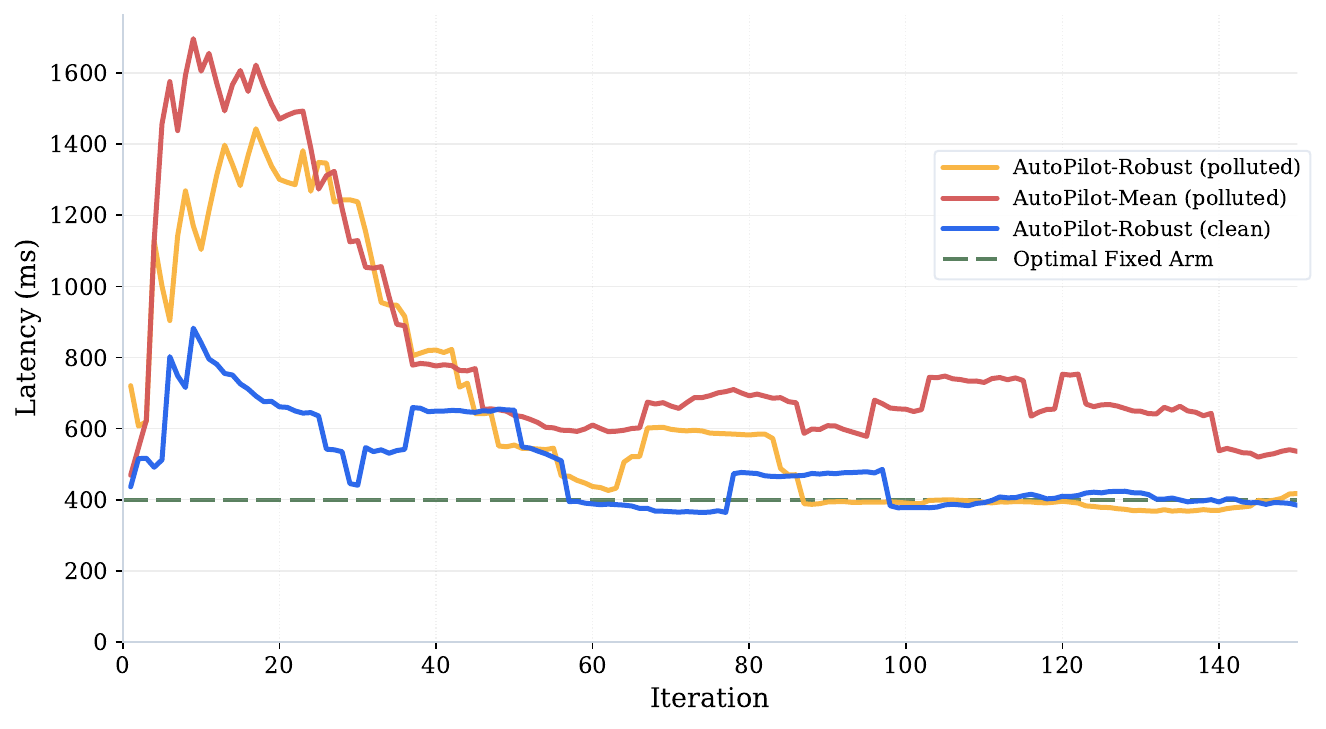}
    \vspace{-2.2em}
    \captionsetup{format=plain, font=scriptsize}
    \caption*{\normalfont \,\,\,\,\, (a) Data Pollution Strategy A}
\end{minipage}
\begin{minipage}{0.4\textwidth}
	\centering
	\includegraphics[width=\columnwidth]{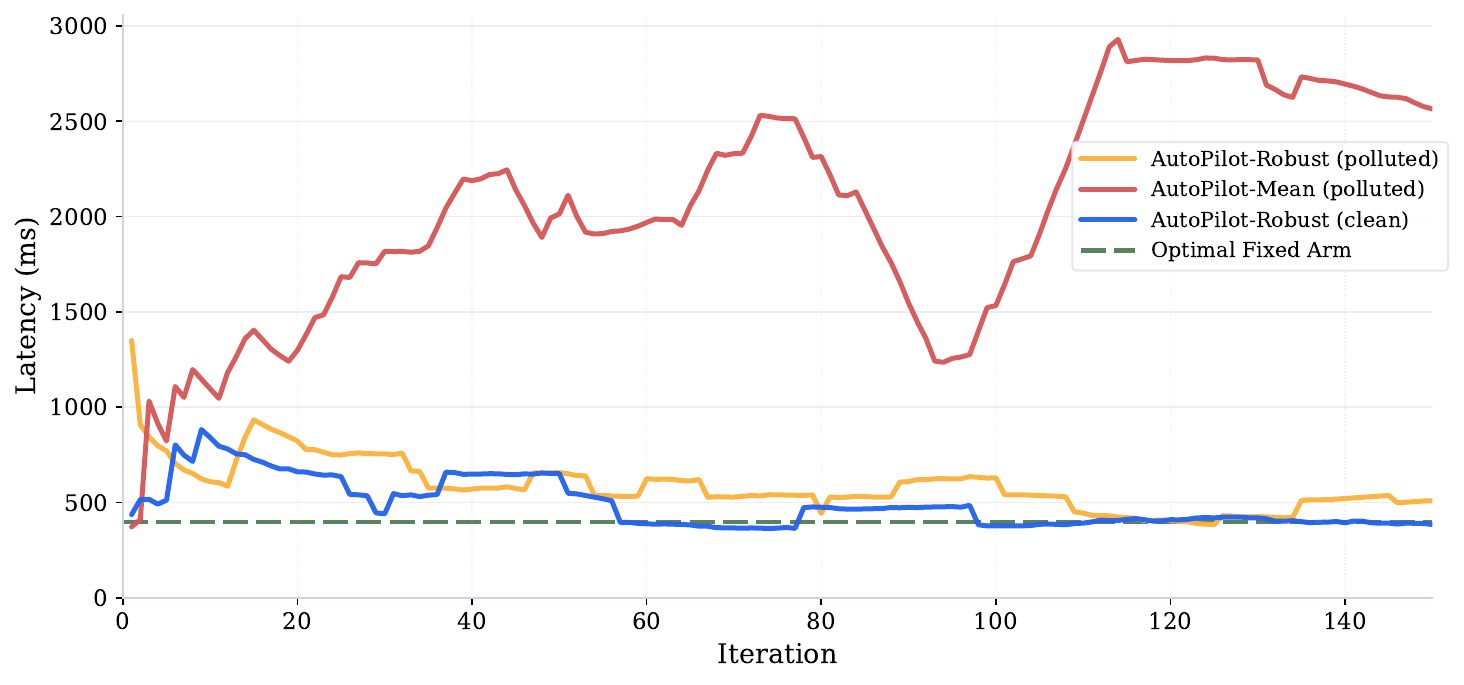}
    \vspace{-2.2em}
    \captionsetup{format=plain, font=scriptsize}
    \caption*{\normalfont \,\,\,\,\, (b) Data Pollution Strategy B}
\end{minipage}
\caption{Robustness of \sys against data pollution}
\vspace{1em}
\label{fig:robust}
\end{figure}

\subsection{Robustness of \sys} \label{subsec:robustness}
We evaluate the robustness of \sys's learning coordination protocol under two adversarial data pollution strategies. The baseline aggregates observations by simply averaging all reported values.

\noindent\uline{Strategy A: Random noise injection.}
Malicious nodes independently scale their observed states and rewards up or down randomly before reporting them to other replicas. This attack injects noisy observations into the distributed training process.

\noindent\uline{Strategy B: Reward manipulation attack.} Malicious nodes randomly scale their observed states as in Strategy A, but craft their reward reports through a carefully designed process. Specifically, each malicious node first estimates the honest nodes' reward reports based on its own local observations, then sets a target aggregated reward $r^*$ to make all actions appear equally effective. Concretely, given $f$ malicious nodes and $2f+1$ honest nodes, the mean aggregate is determined by $\frac{\sum_{h} r_h + \sum_{b} r_m}{3f+1}$, where $r_h$ and $r_m$ denote the honest and malicious reports respectively. Each malicious node computes $\sum_{h} r_h$ using its own local reward measurement as an approximation for honest nodes' reports, then sets $r_b = \frac{(3f+1) \cdot r^* - \sum_{h} r_h}{f}$ such that the resulting mean aggregate equals the target reward $r^*$.

As shown in Figure \ref{fig:robust}(a), the mean aggregation baseline remains partially effective under data pollution strategy A, since the injected noise does not completely destroy the relative ranking between high-performing and low-performing actions. Although the final learned policy is suboptimal, the model can still distinguish better configurations from worse ones. However, as shown in Figure \ref{fig:robust}(b), the impact of strategy B is more severe. Once the learned reward estimates become indistinguishable, unlucky exploration may repeatedly drive the system toward poorly performing actions since the aggregated rewards still appear normal. These results also demonstrate that Byzantine attackers do not necessarily need to report extreme or obviously anomalous values to disrupt learning; instead, carefully crafted yet statistically plausible reports can be significantly more damaging.

Under Strategy A, Byzantine nodes report randomly scaled states and rewards, producing values that deviate significantly from the true observations. Under Strategy B, Byzantine nodes must report a carefully crafted reward value to drive the mean aggregate toward $r^*$: when the true reward falls below $r^*$, the manipulated reward value must be increased beyond the true value, and vice versa. In both cases, the manipulated report values show as the statistical outliers relative to the honest majority. Since \sys's robust aggregation filters out extreme values, both the aggregated state and reward accurately reflect the true observations of the honest majority. Consequently, \sys consistently converges to the optimal configuration under both attack strategies, as shown in Figure~\ref{fig:robust}. Overall, \sys with robust aggregation reduces runtime latency by 27.6\% compared with the baseline under Strategy A, and by 281.8\% under Strategy B.

\vspace{1.5em}
\subsection{Overhead of \sys} \label{subsec:overhead}
Our final experiment evaluates the runtime overhead introduced by \sys. Among all stages in the learning pipeline, only the coordination phase introduces additional network communication in the form of a single all-to-all \textsc{Local\_Report} broadcast at each epoch, as illustrated in Section \ref{sec:coordination}. All the remaining stages, including feature extraction, model inference, and online training, are purely local computations that do not affect inter-node communication.
To evaluate the above costs, we repeat the static experiment from Section~\ref {subsec:static} and profile the overhead of each stage. The results are summarized in Table~\ref{table:overhead} together with the duration of an epoch. 


Among all stages, feature extraction contributes the largest overhead. This is expected since the learning agent continuously computes runtime execution statistics from recent protocol windows, which primarily involves runtime logging and I/O operations.

Communication overhead remains consistently low across all scenarios, mainly since \sys piggybacks coordination data onto Autobahn’s existing consensus pipeline and only introduces a round of broadcast overhead.

Inference overhead is negligible (\(\approx 4\) ms on average), while training overhead remains below \(50\) ms across all scenarios. Unlike deep neural network-based approaches, which often incur substantial online training and inference costs, \sys employs lightweight random forest models with low computational overhead, making online adaptation practical for real-world deployments.

Overall, the overhead introduced by \sys remains modest compared with the duration of each learning epoch. Across all scenarios, the average total overhead is approximately \(0.66\) seconds, accounting for only 4.3\% of the average epoch duration (\(15.05\) seconds). Moreover, since \sys runs asynchronously alongside Autobahn, all stages in the pipeline are performed off the protocol’s critical path and thus introduce no blocking overhead.

\begin{table}[t]
\centering
\caption{Overhead of \sys}
\label{table:overhead}

\scriptsize
\setlength{\tabcolsep}{4pt}

\begin{tabular}{cccccc}
\toprule
\textbf{ } & \textbf{Featurization} & \textbf{Communication} & \textbf{Inference} & \textbf{Training} & \textbf{Epoch} \\
\midrule
\textbf{$S_1$} & 0.371 $\pm$ 0.240s & 0.090 $\pm$ 0.058s & 0.004 $\pm$ 0.001s & 0.044 $\pm$ 0.005s & 6.805 $\pm$ 3.549s \\
\textbf{$S_2$} & 0.496 $\pm$ 0.316s & 0.105 $\pm$ 0.080s & 0.004 $\pm$ 0.001s & 0.042 $\pm$ 0.004s & 15.597 $\pm$ 10.756s \\
\textbf{$S_3$} & 0.674 $\pm$ 0.399s & 0.102 $\pm$ 0.077s & 0.004 $\pm$ 0.001s & 0.043 $\pm$ 0.004s & 22.741 $\pm$ 5.862s \\
\midrule
\textbf{Avg.} & 0.514 $\pm$ 0.318s & 0.099 $\pm$ 0.072s & 0.004 $\pm$ 0.001s & 0.043 $\pm$ 0.004s & 15.048 $\pm$ 6.72s \\
\bottomrule
\end{tabular}
\end{table}

\vspace{0.53em}
\section{Related Work}\label{sec:related_work}

Distributed protocol parameter tuning has attracted increasing attention due to the significant impact of system configurations on throughput and latency. Existing studies mainly focus on automatically optimizing protocol parameters using reinforcement learning or Bayesian optimization techniques. For example, Athena~\cite{li2023auto} proposes a multi-agent DRL framework for tuning permissioned blockchains such as Hyperledger Fabric~\cite{gorenflo2019fastfabric}, while Alzo~\cite{ding2026alzo} further extends this direction to DAG-based protocols using a hierarchical reinforcement learning framework. 
However, most existing works focus on coarse-grained system-level tuning and do not explore fine-grained adaptation of the internal design parameters of consensus protocols under dynamic workloads and fault scenarios.

At the core of \sys, learning in untrusted distributed settings is made resilient against intentional data pollution via robust aggregation and agreement. Similar techniques have been explored in robust distributed learning studies~\cite{allouah2023robust, yin2018byzantine, karimireddy2021learning, farhadkhani2022byzantine, guerraoui2018hidden, baruch2019little}.

More generally, harnessing learning to enhance performance has been done successfully in many systems under the umbrella of machine programming~\cite{pillars}: indexing~\cite{ml_index}, query optimization~\cite{bao, balsa, lero}, concurrency control~\cite{polyjuice}, database tuning~\cite{ml_tuning}, software analysis~\cite{controlflag}, scheduling~\cite{decima}, transaction management in blockchains~\cite{wu2022adachain}.

\section{Conclusion}\label{sec:conc}
Existing BFT protocols lack flexibility and adaptability, leading to suboptimal performance in various scenarios. In this paper, we propose a practical reinforcement learning-based BFT system called \sys, which dynamically selects the top-performing parameter configuration in real-time.
\sys operates without requiring offline profiling or prior data collection. Moreover, its decentralized, Byzantine fault-tolerant coordination protocol ensures that the learning process itself remains resilient to adversarial manipulation.
Our extensive evaluation demonstrates that \sys significantly outperforms existing solutions under various conditions, including dynamic environments and adversarial attacks.
\sys reduces end-to-end latency by 49.8\% compared to the default protocol configuration, and outperforms random configuration exploration by 73.3\%.
While this paper focuses on Autobahn as the state-of-the-art DAG-based BFT protocol, the findings can be generalized to any protocol that combines a DAG-based data dissemination layer and an external leader-based consensus layer (e.g., Narwhal + HotStuff). We believe that \sys establishes a compelling foundation for self-tuning distributed consensus systems, and that the integration of fine-grained learned adaptation into protocol design will become an increasingly important direction as decentralized systems scale to more complex and adversarial operating conditions.

\balance

\bibliographystyle{ACM-Reference-Format}
\bibliography{main}

@article{an2024rlchain,
  title={Rlchain: A drl approach for blockchain performance optimization toward iiot},
  author={An, Min and Zhang, Xuan and Wang, Jishu and Fan, Qiyuan and Gao, Chen and Li, Linyu and Lu, Cuizhen and Li, Nan and Liu, Yingchen},
  journal={IEEE Transactions on Network and Service Management},
  volume={22},
  number={2},
  pages={1629--1645},
  year={2024},
  publisher={IEEE}
}

@inproceedings{ding2026alzo,
  title={Alzo: Auto-Tuning with Reinforcement Learning for DAG-based Blockchains},
  author={Ding, Qiuyu and Zhang, Rongkai and Zhang, Qinnan and Xiao, Zhen and Long, Jieyi and Wan, Mingchao and Liu, Sen and Dong, Jin},
  booktitle={Proceedings of the ACM Web Conference 2026},
  pages={5474--5483},
  year={2026}
}

@article{li2023auto,
  title={Auto-tuning with reinforcement learning for permissioned blockchain systems},
  author={Li, Mingxuan and Wang, Yazhe and Ma, Shuai and Liu, Chao and Huo, Dongdong and Wang, Yu and Xu, Zhen},
  journal={Proceedings of the VLDB Endowment},
  volume={16},
  number={5},
  pages={1000--1012},
  year={2023},
  publisher={VLDB Endowment}
}

@inproceedings{guerraoui2018hidden,
  title={The hidden vulnerability of distributed learning in byzantium},
  author={Guerraoui, Rachid and Rouault, S{\'e}bastien and others},
  booktitle={Int. Conf. on Machine Learning (ICML)},
  pages={3521--3530},
  year={2018},
  organization={PMLR}
}

@inproceedings{yin2018byzantine,
  title={Byzantine-robust distributed learning: Towards optimal statistical rates},
  author={Yin, Dong and Chen, Yudong and Kannan, Ramchandran and Bartlett, Peter},
  booktitle={Int. Confe. on Machine Learning (ICML)},
  pages={5650--5659},
  year={2018}
}

@inproceedings{allouah2023robust,
  title={Robust Distributed Learning: Tight Error Bounds and Breakdown Point under Data Heterogeneity},
  author={Allouah, Youssef and Guerraoui, Rachid and Gupta, Nirupam and Pinot, Rafael and Rizk, Geovani},
  booktitle={Conf. on Neural Information Processing Systems (NeurIPS)},
  year={2023}
}

@inproceedings{pillars,
	location = {Philadelphia, {PA}, {USA}},
	title = {The three pillars of machine programming},
	isbn = {978-1-4503-5834-7},
	url = {https://doi.org/10.1145/3211346.3211355},
	doi = {10.1145/3211346.3211355},
	series = {{MAPL} 2018},
	pages = {69--80},
	booktitle = {Proceedings of the 2nd {ACM} {SIGPLAN} International Workshop on Machine Learning and Programming Languages},
	publisher = {Association for Computing Machinery},
	author = {Gottschlich, Justin and Solar-Lezama, Armando and Tatbul, Nesime and Carbin, Michael and Rinard, Martin and Barzilay, Regina and Amarasinghe, Saman and Tenenbaum, Joshua B. and Mattson, Tim},
	urldate = {2020-02-13},
	date = {2018-06-18},
}

@article{thompson_bootstrap,
	title = {Bootstrapped Thompson Sampling and Deep Exploration},
	url = {http://arxiv.org/abs/1507.00300},
	journaltitle = {{arXiv}:1507.00300 [cs, stat]},
	author = {Osband, Ian and Van Roy, Benjamin},
	urldate = {2020-01-30},
	date = {2015-07-01},
}

@article{bandit_survey,
	title = {A Survey on Contextual Multi-armed Bandits},
	url = {http://arxiv.org/abs/1508.03326},
	journaltitle = {{arXiv}:1508.03326 [cs]},
	author = {Zhou, Li},
	urldate = {2020-01-30},
	date = {2016-02-01},
}

@article{decima,
	title = {Learning Scheduling Algorithms for Data Processing Clusters},
	url = {http://arxiv.org/abs/1810.01963},
	series = {{arXiv} '18},
	journaltitle = {{arXiv}:1810.01963 [cs, stat]},
	author = {Mao, Hongzi and Schwarzkopf, Malte and Venkatakrishnan, Shaileshh Bojja and Meng, Zili and Alizadeh, Mohammad},
	urldate = {2019-06-01},
	date = {2018},
	eprinttype = {arxiv},
	eprint = {1810.01963},
}

@inproceedings{thompson_bound,
	title = {Further Optimal Regret Bounds for Thompson Sampling},
	series = {{AISTATS} '13},
	booktitle = {The International Conference on Artificial Intelligence and Statistics},
	author = {Agrawal, Shipra and Goyal, Navin},
	date = {2013},
}

@inproceedings{bootstrapping,
	title = {Bagging Predictors},
	series = {Maching Learning '96},
	booktitle = {Machine Learning},
	author = {Breiman, Leo},
	date = {1996},
}

@inproceedings{thompson_intro,
	title = {An empirical evaluation of Thompson sampling},
	series = {{NIPS}'11},
	booktitle = {Advances in neural information processing systems},
	author = {Chapelle, Olivier and Li, Lihong},
	date = {2011},
}

@inproceedings{controlflag,
	location = {New York, {NY}, {USA}},
	title = {{ControlFlag}: a self-supervised idiosyncratic pattern detection system for software control structures},
	isbn = {978-1-4503-8467-4},
	url = {https://doi.org/10.1145/3460945.3464954},
	doi = {10.1145/3460945.3464954},
	series = {{MAPS} '21},
	shorttitle = {{ControlFlag}},
	pages = {32--42},
	booktitle = {Proceedings of the 5th {ACM} {SIGPLAN} International Symposium on Machine Programming},
	publisher = {Association for Computing Machinery},
	author = {Hasabnis, Niranjan and Gottschlich, Justin},
	urldate = {2023-03-14},
	date = {2021-06-20},
}

@inproceedings{ml_tuning,
	location = {New York, {NY}, {USA}},
	title = {Automatic Database Management System Tuning Through Large-scale Machine Learning},
	isbn = {978-1-4503-4197-4},
	url = {http://doi.acm.org/10.1145/3035918.3064029},
	doi = {10.1145/3035918.3064029},
	series = {{SIGMOD} '17},
	pages = {1009--1024},
	booktitle = {Proceedings of the 2017 {ACM} International Conference on Management of Data},
	publisher = {{ACM}},
	author = {Van Aken, Dana and Pavlo, Andrew and Gordon, Geoffrey J. and Zhang, Bohan},
	urldate = {2018-02-27},
	date = {2017},
}

@inproceedings{balsa,
	location = {New York, {NY}, {USA}},
	title = {Balsa: Learning a Query Optimizer Without Expert Demonstrations},
	isbn = {978-1-4503-9249-5},
	url = {https://doi.org/10.1145/3514221.3517885},
	doi = {10.1145/3514221.3517885},
	series = {{SIGMOD} '22},
	shorttitle = {Balsa},
	pages = {931--944},
	booktitle = {Proceedings of the 2022 International Conference on Management of Data},
	publisher = {Association for Computing Machinery},
	author = {Yang, Zongheng and Chiang, Wei-Lin and Luan, Sifei and Mittal, Gautam and Luo, Michael and Stoica, Ion},
	urldate = {2022-09-10},
	date = {2022-06-10},
}

@inproceedings{bao,
	location = {China},
	title = {Bao: Making Learned Query Optimization Practical},
	rights = {All rights reserved},
	isbn = {978-1-4503-8343-1},
	doi = {10.1145/3448016.3452838},
	series = {{SIGMOD} '21},
	eventtitle = {{SIGMOD} '21},
	booktitle = {Proceedings of the 2021 International Conference on Management of Data},
	author = {Marcus, Ryan and Negi, Parimarjan and Mao, Hongzi and Tatbul, Nesime and Alizadeh, Mohammad and Kraska, Tim},
	date = {2021-06},
	note = {Award: 'best paper award'},
}

@inproceedings{ml_index,
	location = {New York, {NY}, {USA}},
	title = {The Case for Learned Index Structures},
	isbn = {978-1-4503-4703-7},
	url = {http://doi.acm.org/10.1145/3183713.3196909},
	doi = {10.1145/3183713.3196909},
	series = {{SIGMOD} '18},
	shorttitle = {Learned Index Structures},
	booktitle = {Proceedings of the 2018 International Conference on Management of Data},
	publisher = {{ACM}},
	author = {Kraska, Tim and Beutel, Alex and Chi, Ed H. and Dean, Jeffrey and Polyzotis, Neoklis},
	urldate = {2018-08-21},
	date = {2018},
}

@article{lero,
	title = {Lero: A Learning-to-Rank Query Optimizer},
	volume = {16},
	issn = {2150-8097},
	url = {https://doi.org/10.14778/3583140.3583160},
	doi = {10.14778/3583140.3583160},
	series = {{VLDB} '23},
	shorttitle = {Lero},
	pages = {1466--1479},
	number = {6},
	journaltitle = {Proceedings of the {VLDB} Endowment},
	shortjournal = {Proc. {VLDB} Endow.},
	author = {Zhu, Rong and Chen, Wei and Ding, Bolin and Chen, Xingguang and Pfadler, Andreas and Wu, Ziniu and Zhou, Jingren},
	urldate = {2023-12-05},
	date = {2023-02-01},
}

@inproceedings{polyjuice,
	title = {Polyjuice: \{High-Performance\} Transactions via Learned Concurrency Control},
	isbn = {978-1-939133-22-9},
	url = {https://www.usenix.org/conference/osdi21/presentation/wang-jiachen},
	series = {{OSDI} '21},
	shorttitle = {Polyjuice},
	eventtitle = {15th {USENIX} Symposium on Operating Systems Design and Implementation},
	pages = {198--216},
	author = {Wang, Jiachen and Ding, Ding and Wang, Huan and Christensen, Conrad and Wang, Zhaoguo and Chen, Haibo and Li, Jinyang},
	urldate = {2023-12-05},
	date = {2021},
	langid = {english},
}

@misc{fantom,
  title = {Fantom},
  howpublished = {{https://fantom.foundation/}},
}

@misc{Aptos,
  title = {Aptos},
  howpublished = {{https://aptosnetwork.com/}},
}

@misc{autobahn-codebase,
  title = {Autobahn Codebase},
  howpublished = {{https://github.com/neilgiri/autobahn-artifact}},
}

@misc{ed25519-dalek,
  title={Dalek elliptic curve cryptography},
  howpublished = {{https://github.com/dalek-cryptography/curve25519-dalek/tree/main/ed25519-dalek}},
}

@misc{Sui,
  title = {Sui},
  howpublished = {{https://sui.io/}},
}

@misc{Avalanche,
  title = {Avalanche},
  howpublished = {{https://www.avax.network/}},
}

@misc{seigiga,
  title = {Sei Giga},
  howpublished = {{https://docs.sei.io/learn/sei-giga}},
}

@misc{stable,
  title = {Stable},
  howpublished = {{https://docs.stable.xyz/}},
}

@misc{Somnia,
  title = {Somnia},
  howpublished = {{https://docs.somnia.network}},
}

@misc{microsoftCCF,
  title = {MicrosoftCCF},
  howpublished = {{https://github.com/microsoft/CCF}},
}

@article{resilientdb,
  title={Global-Scale Sustainable Blockchain Fabric},
  author={Apache ResilientDB},
  howpublished = {{https://resilientdb.incubator.apache.org/}},
}

@misc{rocksdb,
  title = {RocksDB, version 0.16.0.},
  howpublished = {{https://rocksdb.org/}},
}

@misc{tcp,
  title = {Tokio, version 1.5.0.},
  howpublished = {{https://tokio.rs/}},
}

@misc{HyperledgerUrsa,
    author = { HyperLedger },
    howpublished = {{https://github.com/hyperledger/ursa}},
    year = { 2019 }
}

@article{amiri2019caper,
  title={{CAPER}: a cross-application permissioned blockchain},
  author={Amiri, Mohammad Javad and Agrawal, Divyakant and El Abbadi, Amr},
  journal={Proc. of the VLDB Endowment},
  volume={12},
  number={11},
  pages={1385--1398},
  year={2019},
  publisher={VLDB Endowment}
}

@inproceedings{amiri2021sharper,
  title={Shar{P}er: Sharding Permissioned Blockchains Over Network Clusters},
  author={Amiri, Mohammad Javad and Agrawal, Divyakant and El Abbadi, Amr},
  booktitle={SIGMOD Int. Conf. on Management of Data},
  pages={76--88},
  year={2021},
  organization={ACM}
}

@inproceedings{gueta2019sbft,
  title={SBFT: a Scalable Decentralized Trust Infrastructure for Blockchains},
  author={Gueta, Guy Golan and Abraham, Ittai and Grossman, Shelly and Malkhi, Dahlia and Pinkas, Benny and Reiter, Michael K and Seredinschi, Dragos-Adrian and Tamir, Orr and Tomescu, Alin},
  booktitle={Int. Conf. on Dependable Systems and Networks (DSN)},
  pages={568-580},
  year={2019},
  publisher={IEEE/IFIP}
}

@inproceedings{kogias2016enhancing,
  title={Enhancing bitcoin security and performance with strong consistency via collective signing},
  author={Kogias, Eleftherios Kokoris and Jovanovic, Philipp and Gailly, Nicolas and Khoffi, Ismail and Gasser, Linus and Ford, Bryan},
  booktitle={Security Symposium},
  pages={279--296},
  year={2016},
  organization={USENIX Association}
}

@inproceedings{gorenflo2019fastfabric,
  title={Fastfabric: Scaling hyperledger fabric to 20,000 transactions per second},
  author={Gorenflo, Christian and Lee, Stephen and Golab, Lukasz and Keshav, Srinivasan},
  booktitle={Int. Conf. on Blockchain and Cryptocurrency (ICBC)},
  pages={455--463},
  year={2019},
  organization={IEEE}
}

@inproceedings{luu2016secure,
  title={A secure sharding protocol for open blockchains},
  author={Luu, Loi and Narayanan, Viswesh and Zheng, Chaodong and Baweja, Kunal and Gilbert, Seth and Saxena, Prateek},
  booktitle={SIGSAC Conf. on Computer and Communications Security (CCS)},
  pages={17--30},
  year={2016},
  organization={ACM}
}

@inproceedings{kokoris2018omniledger,
  title={Omniledger: A secure, scale-out, decentralized ledger via sharding},
  author={Kokoris-Kogias, Eleftherios and Jovanovic, Philipp and Gasser, Linus and Gailly, Nicolas and Syta, Ewa and Ford, Bryan},
  booktitle={Symposium on Security and Privacy (SP)},
  pages={583--598},
  year={2018},
  organization={IEEE}
}

@article{wu2022adachain,
    title = {Ada{C}hain: A Learned Adaptive Blockchain},
    author = {Wu, Chenyuan and Mehta, Bhavana and Amiri, Mohammad Javad and Marcus, Ryan and Loo, Boon Thau},
    year = {2023},
    publisher = {VLDB Endowment},
    volume = {16},
    number = {8},
    journal={Proc. of the VLDB Endowment},
    pages = {2033–2046}
}

@article{wu2024towards,
  title={Towards Full Stack Adaptivity in Permissioned Blockchains},
  author={Wu, Chenyuan and Amiri, Mohammad Javad and Qin, Haoyun and Mehta, Bhavana and Marcus, Ryan and Loo, Boon Thau},
    year = {2024},
    publisher = {VLDB Endowment},
    volume = {17},
    number = {5},
    journal={Proc. of the VLDB Endowment},
    pages = {1073–1080}
}

@article{brown2016corda,
  title={Corda: an introduction},
  author={Brown, Richard Gendal and Carlyle, James and Grigg, Ian and Hearn, Mike},
  journal={R3 CEV, August},
  volume={1},
  number={15},
  pages={14},
  year={2016}
}

@article{buchnik2020fireledger,
  title={FireLedger: a high throughput blockchain consensus protocol},
  author={Buchnik, Yehonatan and Friedman, Roy},
  journal={Proceedings of the VLDB Endowment},
  volume={13},
  number={9},
  pages={1525--1539},
  year={2020},
  publisher={VLDB Endowment}
}

@article{chacko2023optimize,
  title={How to optimize my blockchain? a multi-level recommendation approach},
  author={Chacko, Jeeta Ann and Mayer, Ruben and Jacobsen, Hans-Arno},
  booktitle={SIGMOD Int. Conf. on Management of Data},
  volume={1},
  number={1},
  pages={1--27},
  year={2023},
  publisher={ACM}
}

@misc{narwhalrepo2026,
  author={{facebookresearch}},
  title={Narwhal},
  year={2026},
  howpublished={\url{https://github.com/facebookresearch/narwhal}},
  note={Public code repository for Narwhal and Tusk}
}

@article{breiman2001random,
  title={Random forests},
  author={Breiman, Leo},
  journal={Machine learning},
  volume={45},
  pages={5--32},
  year={2001},
  publisher={Springer}
}

@article{baruch2019little,
  title={A little is enough: Circumventing defenses for distributed learning},
  author={Baruch, Gilad and Baruch, Moran and Goldberg, Yoav},
  journal={Advances in Neural Information Processing Systems},
  volume={32},
  year={2019}
}

@inproceedings{farhadkhani2022byzantine,
  title={Byzantine machine learning made easy by resilient averaging of momentums},
  author={Farhadkhani, Sadegh and Guerraoui, Rachid and Gupta, Nirupam and Pinot, Rafael and Stephan, John},
  booktitle={Int. Conf. on Machine Learning (ICML)},
  pages={6246--6283},
  year={2022},
  organization={PMLR}
}

@inproceedings{karimireddy2021learning,
  title={Learning from history for byzantine robust optimization},
  author={Karimireddy, Sai Praneeth and He, Lie and Jaggi, Martin},
  booktitle={Int. Conf. on Machine Learning (ICML)},
  pages={5311--5319},
  year={2021},
  organization={PMLR}
}

@article{bessani2013depsky,
  title={DepSky: dependable and secure storage in a cloud-of-clouds},
  author={Bessani, Alysson and Correia, Miguel and Quaresma, Bruno and Andr{\'e}, Fernando and Sousa, Paulo},
  journal={Transactions on Storage (TOS)},
  volume={9},
  number={4},
  pages={12},
  year={2013},
  publisher={ACM}
}

@inproceedings{clement2009upright,
  title={Upright cluster services},
  author={Clement, Allen and Kapritsos, Manos and Lee, Sangmin and Wang, Yang and Alvisi, Lorenzo and Dahlin, Mike and Riche, Taylor},
  booktitle={Symposium on Operating Systems Principles (SOSP)},
  pages={277--290},
  year={2009},
  organization={ACM}
}

@inproceedings{nagda2026dag,
  title={DAG of DAGs: Order-Fairness Made Practical},
  author={Nagda, Heena and  Sankhe, Sidharth and Sinha, Sakshi and Attarha, Keon and Amiri, Mohammad Javad and Loo, Boon Thau},
  booktitle={SIGMOD Int. Conf. on Management of Data},
  year={2026},
  organization={ACM}
}

@inproceedings{castro1999practical,
  title={Practical Byzantine fault tolerance},
  author={Castro, Miguel and Liskov, Barbara},
  booktitle={Symposium on Operating Systems Design and Implementation (OSDI)},
  pages={173--186},
  year={1999},
  organization={USENIX Association}
}

@article{castro2002practical,
  title={Practical Byzantine fault tolerance and proactive recovery},
  author={Castro, Miguel and Liskov, Barbara},
  journal={Transactions on Computer Systems (TOCS)},
  volume={20},
  number={4},
  pages={398--461},
  year={2002},
  publisher={ACM}
}

@article{kotla2007zyzzyva,
  title={Zyzzyva: speculative byzantine fault tolerance},
  author={Kotla, Ramakrishna and Alvisi, Lorenzo and Dahlin, Mike and Clement, Allen and Wong, Edmund},
  journal={Operating Systems Review (OSR)},
  volume={41},
  number={6},
  pages={45--58},
  year={2007},
  publisher={ACM SIGOPS}
}

@inproceedings{clement2009making,
  title={Making Byzantine Fault Tolerant Systems Tolerate Byzantine Faults.},
  author={Clement, Allen and Wong, Edmund L and Alvisi, Lorenzo and Dahlin, Michael and Marchetti, Mirco},
  booktitle={Symposium on Networked Systems Design and Implementation (NSDI)},
  volume={9},
  pages={153--168},
  year={2009},
  organization={USENIX Association}
}

@inproceedings{yin2019hotstuff,
  title={HotStuff: BFT consensus with linearity and responsiveness},
  author={Yin, Maofan and Malkhi, Dahlia and Reiter, Michael K and Gueta, Guy Golan and Abraham, Ittai},
  booktitle={Symposium on Principles of Distributed Computing (PODC)},
  pages={347--356},
  year={2019},
  organization={ACM}
}

@article{amir2011prime,
  title={Prime: Byzantine replication under attack},
  author={Amir, Yair and Coan, Brian and Kirsch, Jonathan and Lane, John},
  journal={Transactions on Dependable and Secure Computing},
  volume={8},
  number={4},
  pages={564--577},
  year={2011},
  publisher={IEEE}
}

@article{kieckhafer1994reaching,
  title={Reaching approximate agreement with mixed-mode faults},
  author={Kieckhafer, Roger M. and Azadmanesh, Mohammad H.},
  journal={Transactions on Parallel and Distributed Systems},
  volume={5},
  number={1},
  pages={53--63},
  year={1994},
  publisher={IEEE}
}

@article{dwork1988consensus,
  title={Consensus in the presence of partial synchrony},
  author={Dwork, Cynthia and Lynch, Nancy and Stockmeyer, Larry},
  journal={Journal of the ACM (JACM)},
  volume={35},
  number={2},
  pages={288--323},
  year={1988},
  publisher={ACM}
}

@article{garcia2016sieveq,
  title={SieveQ: A layered bft protection system for critical services},
  author={Garcia, Miguel and Neves, Nuno and Bessani, Alysson},
  journal={IEEE Transactions on Dependable and Secure Computing},
  volume={15},
  number={3},
  pages={511--525},
  year={2016},
  publisher={IEEE}
}

@inproceedings{babay2019deploying,
  title={Deploying intrusion-tolerant SCADA for the power grid},
  author={Babay, Amy and Schultz, John and Tantillo, Thomas and Beckley, Samuel and Jordan, Eamon and Ruddell, Kevin and Jordan, Kevin and Amir, Yair},
  booktitle={Int. Conf. on Dependable Systems and Networks (DSN)},
  pages={328--335},
  year={2019},
  organization={IEEE}
}

@inproceedings{adya2002farsite,
  title={${FARSITE}$: Federated, Available, and Reliable Storage for an Incompletely Trusted Environment},
  author={Adya, Atul and Bolosky, William J and Castro, Miguel and Cermak, Gerald and Chaiken, Ronnie and Douceur, John R and Howell, Jon and Lorch, Jacob R and Theimer, Marvin and Wattenhofer, Roger P},
  booktitle={Symposium on Operating Systems Design and Implementation (OSDI)},
  year={2002},
 organization={USENIX Association}
}

@article{sousa2009highly,
  title={Highly available intrusion-tolerant services with proactive-reactive recovery},
  author={Sousa, Paulo and Bessani, Alysson Neves and Correia, Miguel and Neves, Nuno Ferreira and Verissimo, Paulo},
  journal={IEEE Transactions on Parallel and Distributed Systems},
  volume={21},
  number={4},
  pages={452--465},
  year={2009},
  publisher={IEEE}
}

@article{roeder2010proactive,
  title={Proactive obfuscation},
  author={Roeder, Tom and Schneider, Fred B},
  journal={ACM Transactions on Computer Systems (TOCS)},
  volume={28},
  number={2},
  pages={1--54},
  year={2010},
  publisher={ACM}
}

@inproceedings{malkhi1998secure,
  title={Secure and scalable replication in Phalanx},
  author={Malkhi, Dahlia and Reiter, Michael K},
  booktitle={Symposium on Reliable Distributed Systems (SRDS)},
  pages={51--58},
  year={1998},
  organization={IEEE}
}

@article{zhou2002coca,
  title={COCA: A secure distributed online certification authority},
  author={Zhou, Lidong and Schneider, Fred B and Van Renesse, Robbert},
  journal={ACM Transactions on Computer Systems (TOCS)},
  volume={20},
  number={4},
  pages={329--368},
  year={2002},
  publisher={ACM}
}

@inproceedings{dobre2013powerstore,
  title={PoWerStore: Proofs of writing for efficient and robust storage},
  author={Dobre, Dan and Karame, Ghassan and Li, Wenting and Majuntke, Matthias and Suri, Neeraj and Vukoli{\'c}, Marko},
  booktitle={Conf. on Computer and communications security (CCS)},
  pages={285--298},
  year={2013},
  organization={ACM}
}

@inproceedings{goodson2004efficient,
  title={Efficient Byzantine-tolerant erasure-coded storage},
  author={Goodson, Garth R and Wylie, Jay J and Ganger, Gregory R and Reiter, Michael K},
  booktitle={Int. Conf. on Dependable Systems and Networks (DSN)},
  pages={135--144},
  year={2004},
  organization={IEEE}
}

@inproceedings{danezis2022narwhal,
  title={Narwhal and Tusk: a DAG-based mempool and efficient BFT consensus},
  author={Danezis, George and Kokoris-Kogias, Lefteris and Sonnino, Alberto and Spiegelman, Alexander},
  booktitle={European Conf. on Computer Systems (EuroSys)},
  pages={34--50},
  year={2022}
}

@inproceedings{keidar2021all,
  title={All you need is dag},
  author={Keidar, Idit and Kokoris-Kogias, Eleftherios and Naor, Oded and Spiegelman, Alexander},
  booktitle={Symposium on Principles of Distributed Computing (PODC)},
  pages={165--175},
  year={2021},
  publisher={ACM}
}

@inproceedings{spiegelman2022bullshark,
  title={Bullshark: Dag bft protocols made practical},
  author={Spiegelman, Alexander and Giridharan, Neil and Sonnino, Alberto and Kokoris-Kogias, Lefteris},
  booktitle={ACM SIGSAC Conf. on Computer and Communications Security (CCS)},
  pages={2705--2718},
  year={2022}
}

@inproceedings{bahsoun2015making,
  title={Making BFT protocols really adaptive},
  author={Bahsoun, Jean-Paul and Guerraoui, Rachid and Shoker, Ali},
  booktitle={Int. Parallel and Distributed Processing Symposium},
  pages={904--913},
  year={2015},
  organization={IEEE}
}

@inproceedings{wu2025bftbrain,
  title={$\{$BFTBrain$\}$: Adaptive $\{$BFT$\}$ Consensus with Reinforcement Learning},
  author={Wu, Chenyuan and Qin, Haoyun and Amiri, Mohammad Javad and Loo, Boon Thau and Malkhi, Dahlia and Marcus, Ryan},
  booktitle={22nd USENIX Symposium on Networked Systems Design and Implementation (NSDI 25)},
  pages={1563--1583},
  year={2025}
}

@inproceedings{dai2024lightdag,
  title={LightDAG: A low-latency DAG-based BFT consensus through lightweight broadcast},
  author={Dai, Xiaohai and Wang, Guanxiong and Xiao, Jiang and Guo, Zhengxuan and Hao, Rui and Xie, Xia and Jin, Hai},
  booktitle={Int. Parallel and Distributed Processing Symposium (IPDPS)},
  pages={998--1008},
  year={2024},
  organization={IEEE}
}

@inproceedings{dai2023gradeddag,
  title={GradedDAG: An asynchronous DAG-based BFT consensus with lower latency},
  author={Dai, Xiaohai and Zhang, Zhaonan and Xiao, Jiang and Yue, Jingtao and Xie, Xia and Jin, Hai},
  booktitle={Int. Symposium on Reliable Distributed Systems (SRDS)},
  pages={107--117},
  year={2023},
  organization={IEEE}
}

@inproceedings{spiegelman2024shoal,
  title={Shoal: Improving dag-bft latency and robustness},
  author={Spiegelman, Alexander and Arun, Balaji and Gelashvili, Rati and Li, Zekun},
  booktitle={Int. Conf. on Financial Cryptography and Data Security (FC)},
  pages={92--109},
  year={2024},
  organization={Springer}
}

@inproceedings{arun2025shoal++,
  title={Shoal++: High Throughput DAG BFT Can Be Fast and Robust!},
  author={Arun, Balaji and Li, Zekun and Suri-Payer, Florian and Das, Sourav and Spiegelman, Alexander},
  booktitle={Symposium on Networked Systems Design and Implementation (NSDI)},
  year={2025},
  organization={USENIX Association}
}

@inproceedings{malkhi2024bbca,
  title={BBCA-CHAIN: Low latency, high throughput BFT consensus on a DAG},
  author={Malkhi, Dahlia and Stathakopoulou, Chrysoula and Yin, Maofan},
  booktitle={Int. Conf. on Financial Cryptography and Data Security (FC)},
  pages={51--73},
  year={2024},
  organization={Springer}
}

@article{dai2024remora,
  title={Remora: A low-latency dag-based bft through optimistic paths},
  author={Dai, Xiaohai and Li, Wei and Wang, Guanxiong and Xiao, Jiang and Chen, Haoyang and Li, Shufei and Zomaya, Albert Y and Jin, Hai},
  journal={Transactions on Computers},
  year={2024},
  publisher={IEEE}
}

@article{dai2024wahoo,
  title={Wahoo: A dag-based bft consensus with low latency and low communication overhead},
  author={Dai, Xiaohai and Zhang, Zhaonan and Guo, Zhengxuan and Ding, Chaozheng and Xiao, Jiang and Xie, Xia and Hao, Rui and Jin, Hai},
  journal={Transactions on Information Forensics and Security},
  year={2024},
  publisher={IEEE}
}

@inproceedings{shrestha2024sailfish,
  title={Sailfish: Towards Improving the Latency of DAG-based BFT},
  author={Shrestha, Nibesh and Shrothrium, Rohan and Kate, Aniket and Nayak, Kartik},
  booktitle={Symposium on Security and Privacy (SP)},
  pages={21--21},
  year={2024},
  organization={IEEE}
}

@inproceedings{cheng2024shardag,
  title={Shardag: Scaling dag-based blockchains via adaptive sharding},
  author={Cheng, Feng and Xiao, Jiang and Liu, Cunyang and Zhang, Shijie and Zhou, Yifan and Li, Bo and Li, Baochun and Jin, Hai},
  booktitle={Int. Conf. on Data Engineering (ICDE)},
  pages={2068--2081},
  year={2024},
  organization={IEEE}
}

@inproceedings{giridharan2024autobahn,
  title={Autobahn: Seamless high speed BFT},
  author={Giridharan, Neil and Suri-Payer, Florian and Abraham, Ittai and Alvisi, Lorenzo and Crooks, Natacha},
  booktitle={Symposium on Operating Systems Principles (SOSP)},
  pages={1--23},
  year={2024},
  organization={ACM SIGOPS}
}

@article{kang2025hotstuff,
  title={Hotstuff-1: Linear consensus with one-phase speculation},
  author={Kang, Dakai and Gupta, Suyash and Malkhi, Dahlia and Sadoghi, Mohammad},
  journal={Proceedings of the ACM on Management of Data},
  volume={3},
  number={3},
  pages={1--29},
  year={2025},
  publisher={ACM}
}

@article{kang2025fairdag,
  title={FairDAG: consensus fairness over multi-proposer causal design},
  author={Kang, Dakai and Chen, Junchao and Dinh, Tien Tuan Anh and Sadoghi, Mohammad},
  journal={Proceedings of the VLDB Endowment},
  volume={19},
  number={2},
  pages={265--278},
  year={2025},
  publisher={VLDB Endowment}
}

@article{li2025flexim,
  title={FlexIM: efficient and verifiable index management in Blockchain},
  author={Li, Binhong and Lin, Licheng and Zhang, Shijie and Xu, Jianliang and Xiao, Jiang and Li, Bo and Jin, Hai},
  journal={IEEE Transactions on Knowledge and Data Engineering},
  year={2025},
  publisher={IEEE}
}

\end{document}